%% file: paper.tex
\documentclass[useAMS,usenatbib]{mn2e}

\usepackage{times,mathptmx,graphicx}
\usepackage{aas_macros}
\usepackage{amsmath}
\usepackage{rotating}
\usepackage{url}

\newcommand{\GODUNOV}{Shock-Capturing}
\newcommand{\Godunov}{Shock-capturing}
\newcommand{\godunov}{shock-capturing}

\newcommand{\PENCILLIKE}{High-Order Finite-Difference}

\newcommand{\pencillike}{high-order finite-difference}

\newcommand{\UPWIND}{Upwind}
\newcommand{\Upwind}{Upwind}
\newcommand{\upwind}{upwind}

\newcommand{\zeus}{{\sc ZEUS}}
\newcommand{\nirvana}{{\sc Nirvana}}
\newcommand{\flash}{{\sc FLASH}}

\newcommand{\dangelo}{{\sc Nirvana-GDA}}
\newcommand{\gerben}{{\sc Nirvana-GD}}
\newcommand{\cresswell}{\textsc{Nirvana-PC}}
\newcommand{\kley}{{\sc RH2D}}
\newcommand{\fromang}{{\sc GLOBAL}}
\newcommand{\fargo}{{\sc FARGO}}
\newcommand{\pierens}{{\sc GENESIS}}
\newcommand{\trampvanleer}{{\sc TRAMP-vanLeer}}
\newcommand{\pencil}{{\sc Pencil}}
\newcommand{\pci}{{\sc AMRA}}
\newcommand{\artur}{{\sc FLASH-AG}}
\newcommand{\adam}{{\sc FLASH-AP}}
\newcommand{\sijme}{{\sc Rodeo}}
\newcommand{\jupiter}{{\sc JUPITER}}
\newcommand{\trampppm}{{\sc TRAMP-PPM}}
\newcommand{\treesph}{{\sc SPHTREE}}
\newcommand{\parasph}{{\sc PARASPH}}

\newcommand{\mvect}[1]{\mathbf{#1}}
\newcommand{\grad}[1]{\nabla #1}

\usepackage{hyperref}
\usepackage{thumbpdf}

\title{A comparative study of disc-planet interaction}

\author[M. de Val-Borro et al.]{M.~de~Val-Borro$^1$\thanks{Email: miguel@astro.su.se},
R.~G.~Edgar$^{1,2}$,
P.~Artymowicz$^{1,3}$,
P.~Ciecielag$^{4,5}$, 
P.~Cresswell$^6$,
\newauthor
G.~D'Angelo$^7$,
E.~J.~Delgado-Donate$^1$,
G.~Dirksen$^8$,
S.~Fromang$^{6,9}$,
A.~Gawryszczak$^5$,
\newauthor
H.~Klahr$^{10}$,
W.~Kley$^8$,
W.~Lyra$^{11}$,
F.~Masset$^{12,13}$,
G.~Mellema$^{14,1}$,
R.~P.~Nelson$^6$,
\newauthor
S.-J.~Paardekooper$^{14}$,
A.~Peplinski$^1$,
A.~Pierens$^{15,6}$,
T.~Plewa$^{16}$,
K.~Rice$^{17}$,
C.~Sch\"{a}fer$^8$,
\newauthor
R.~Speith$^8$\\
$^1$ Stockholm University, AlbaNova University Center, SE-106 91, Stockholm, Sweden\\
$^2$ Dept. of Physics and Astronomy, University of Rochester, NY 14627, USA\\
$^3$ University of Toronto at Scarborough, 1265 Military Trail, Toronto, Ontario M1C 1A4, Canada\\
$^4$ University Observatory Munich, Scheinerstr. 1, D-81679 Munich, Germany \\
$^5$ Nicolaus Copernicus Astronomical Centre, Bartycka 18, Warsaw, PL-00-716, Poland\\
$^6$ Astronomy Unit, Queen Mary, University of London, Mile End Rd, London E1 4NS, UK\\
$^7$ School of Physics, University of Exeter, Stocker Road, Exeter, EX4 4QL, UK\\
$^8$ Institute of Astronomy and Astrophysics T\"{u}bingen, Auf der Morgenstelle 10, D-72076 T\"{u}bingen, Germany\\
$^9$ DAMTP, University of
Cambridge, Centre for Mathematical Sciences, Wilberforce Road, Cambridge, CB3 0WA, UK\\
$^{10}$ Max-Planck-Institut f\"{u}r Astronomie, K\"{o}nigstuhl 17, D-69117 Heidelberg, Germany\\
$^{11}$ Department of Astronomy \& Space Physics, Uppsala Astronomical Observatory, Box 515, 751 20, Sweden\\
$^{12}$ CEA, Service d'Astrophysique, Saclay, 91191 Gif-sur-Yvette Cedex, France\\
$^{13}$ IA-UNAM, Ciudad Universitaria, Apartado Postal 70-264, Mexico D.F. 04510, Mexico\\
$^{14}$ Leiden Observatory, P.O. Box 9513, NL-2300 RA Leiden, The Netherlands\\
$^{15}$ Luth, Observatoire de Paris-Meudon, 92 195 Meudon Cedex, France\\
$^{16}$ ASC FLASH Center,  University of Chicago, 5640 South Ellis, Chicago, IL 60637, USA\\
$^{17}$ Scottish Universities Physics Alliance, Institute for Astronomy, University of Edinburgh, Blackford Hill, Edinburgh, EH9 3HJ, UK}

\begin{document}

\date{\today}

\pagerange{\pageref{firstpage}--\pageref{lastpage}} \pubyear{2005}

\maketitle

\label{firstpage}

\begin{abstract}
We perform numerical simulations of a disc-planet system
using various grid-based and
smoothed particle hydrodynamics (SPH) codes.
The tests are run for a simple setup where
Jupiter and Neptune mass planets on a circular orbit
open a gap in a protoplanetary disc
during a few hundred orbital periods.
We compare the surface density contours, potential vorticity
and smoothed radial profiles at several times.
The disc mass and gravitational torque time evolution
are analyzed with high temporal resolution.
There is overall
consistency between the codes.
The density profiles agree within about 5\% for the Eulerian
simulations while the SPH results predict the correct
shape of the gap
although have
less resolution in the low density regions
and weaker planetary wakes.
The disc masses after 200 orbital periods agree within 10\%.
The spread is larger in the tidal torques acting on the planet
which agree within a factor 2 at the end of the simulation.
In the Neptune case the dispersion in the torques is greater 
than for Jupiter,
possibly owing to the contribution from the not completely 
cleared region close to the planet.
\end{abstract}

\begin{keywords}
Physical data and processes: accretion, accretion discs -- hydrodynamics.
Solar system: planets and satellites: general.
\end{keywords}


\input{intro2}


\input{introduction}


\input{description}


\input{codes2}



\input{results}




\input{discussion}



\input{comparison.bbl}

\input{advice.tex}

\section*{Acknowledgments}

This project is supported by the European Research Training Network
``The Origin of Planetary Systems''
(contract number HPRN-CT-2002-00308).
The participants had the opportunity to discuss the
results of the comparison at a workshop in Stockholm in May 2004
funded by the Network.

MdVB is supported by a NOT/IAC scholarship.
RGE, PC (Munich) and GD acknowledge financial support provided through the European Community's Human Potential Programme under contract HPRN-CT-2002-00308, PLANETS.
PC (QMUL) is supported by a PPARC PhD studentship.
GDA was supported by the Leverhulme Trust through a UKAFF Fellowship.
AG is supported through grant 1P03D 02626 from the Polish Ministry of Science.
This work is supported in part by the U.S. Department of Energy under
Grant No.\ B523820 to the Center for Astrophysical Thermonuclear
Flashes at the University of Chicago.
\artur{} calculations were performed at the
Interdisciplinary Centre for Mathematical and Computational
Modeling in Warsaw, Poland.

This research has made use of NASA's Astrophysics Data System Bibliographic Services.

\label{lastpage}

\bsp

\end{document}

%% file: intro2.tex
\section{Introduction}

Hydrodynamics is a difficult subject, which has caused many problems for many
distinguished physicists. However, it is not a topic which can be avoided, due
to the central part that gas plays in the cosmos.

The basic equations of hydrodynamics are the Navier-Stokes equations,
and have been known for almost two centuries:
\begin{eqnarray}
\frac{\partial\rho}{\partial t} + \nabla \cdot (\rho \mvect{v}) & = & 0 \\
\frac{\partial\mvect{v}}{\partial t} + (\mvect{v} \cdot \nabla ) \mvect{v} & = & - \frac{1}{\rho}\grad{p} - \grad{\Phi} + \nabla \cdot {\mathbf T}
\label{eq:NS}
\end{eqnarray}
where $\rho$ is the density,
${\mathbf v}$ the velocity of the fluid,
$p$ the pressure,
$\Phi$ the gravitational potential
and ${\mathbf T}$ is the full viscous stress tensor
\citep[see e.g.][]{1984frh..book.....M}. 
The first equation describes the conservation of mass and the second, conservation of momentum.
An equation of state closes the system of equations, and additional terms
may be added as required.
Despite their comparatively simple form, the Navier-Stokes equations have proved remarkably stubborn to mathematical analysis.

The problem lies in the  $\mathbf{v}\cdot \nabla \mathbf{v}$ terms (the so-called `advection' terms).
These arise because the equations describe a fluid moving past a fixed point in space (the Eulerian point of view).
The advection terms make the equations non-linear (since they are effectively proportional to $v^{2}$), rendering many mathematical techniques useless.
Indeed, no one has yet proven that solutions to the Navier-Stokes equations are unique.
This is in sharp contrast to many other important equations in physics.
For example, the Poisson equation is linear, and has unique solutions.
This opens up many avenues for obtaining solutions --- the Method of Images being a well known example.
As a result of this non-linearity, theoretical investigation of fluids has to be restricted to highly idealised flows.

To make progress then, we are forced to turn to computers.
Numerical algorithms for solving complex equations have been studied for centuries, and computers are ideal for implementing these.
Unfortunately, computers are tricky beasts, with a habit of doing precisely what you told them to, just when you least expected it.
The `obvious' way of computing a numerical solution may well be unstable (this is particularly true of the Navier-Stokes equations), and an implementation of a stable method may well contain bugs.
Floating point numbers have finite accuracy, and various subtleties arise when codes approach this limit.
In particular, arithmetic ceases to be distributive and tests for equality cease to be reliable.
Different architectures, operating systems and compilers all add to the mix.
For this reason, work performed on computers is better described as a `numerical experiment' rather than a `simulation.'

Fortunately, there is no need to be overly pessimistic about the situation.
For example, although the minutiae of floating point numbers and the vagaries of different compilers can be troublesome, these should not give problems in the majority of cases.
Unless agreement to the last bit is required, there should be no significant difference in results obtained with different architectures and compilers.
Instead, it suffices to focus on differences between algorithms for solving the Navier-Stokes equations.
Of course, all hydrodynamics codes are carefully tested against simple problems (such as shock-tubes).
It is on more complex problems that differences and difficulties can be exposed.

Within the context of the EU-RTN ``The Origin of Planetary Systems,''
\footnote{http://www.usm.uni-muenchen.de/Planets/}
we have conducted a comparison of hydrodynamics codes, which we present in this paper.
The problem we selected was that of a planet in a fixed circular orbit in a circumstellar disc.
This has the virtue of simplicity, while still retaining sufficient complexity to allow us to see meaningful differences between the various algorithms.
We ran the test problem on 17 independent codes.

A comparison of several numerical methods on the problem of a planet
embedded in a disc was performed by \citet{1999ApJ...514..344B} using
SPH, van Leer and Godunov methods with different equations of state.
In particular, they studied the accretion onto the planet after
it had cleared a gap.
Other examples of comparisons in different fields to verify algorithms
and implementations
published during the last few years 
include the Santa Barbara cluster project \citep{1999ApJ...525..554F},
the non-LTE radiative transfer code comparison \citep{2002A&A...395..373V},
the Rayleigh-Taylor instability study by
the Alpha-Group collaboration \citep{2004PhFl...16.1668D}
and the comparison of models of
photoionization regions \citep{2001ASPC..247..533P}.

The aim of this project is to test the reliability of
present numerical computations of disc-planet interaction
with a quantitative comparison
and generate a benchmark for future simulations.
In Section~\ref{sec:theory}, we briefly describe the
interaction between a planet and a protoplanetary disc
and outline the motivation for this study.
The initial setup and boundary conditions of the problem are
described in Section~\ref{sec:setup}.
In Section~\ref{sec:codedescribe},
the numerical methods used in the comparison are described.
The results are shown in Section~\ref{sec:results}.
We discuss the results
in Section~\ref{sec:discussion},
and in Appendix~\ref{sec:appendix}
we summarize our experience
with this project that could be useful for future comparisons.

%% file: introduction.tex

\section{Disc-planet interaction}
\label{sec:theory}

Over 160 extrasolar planetary systems have been discovered
by radial velocity measurements
during the last years
\citep[e.g.][]{1995Natur.378..355M,1996ApJ...464L.147M}.
Giant planets have been found in very close orbits
around the central star with orbital periods of a few days
and almost circular orbits,
the so-called ``Hot Jupiters''.
Planets orbiting at larger distances from the star
show a broad eccentricity
distribution 
reaching roughly $e = 0.9$
\citep[for recent reviews of the properties of
the observed systems see][]{2003ASPC..294....1M,Marcy2005}. 
The origin of the differences with the
planets in the solar system
is not well understood, although
various explanations have been proposed.
The standard models explain giant planet formation either
through planetesimal accumulation followed by rapid gas
accretion onto the planet core \citep{1996Icar..124...62P}
or gravitational instabilities in the disc
\citep[see e.g.][]{1998ApJ...503..923B,2001ApJ...563..367B}.
In both cases the planets are likely to have formed at
larger distances from the central star than observed.

Orbital migration due to gravitational
interaction between the planet and the gaseous disc is a possible
mechanism to bring planets to a close orbit.
The tidal interaction between a planet and a
gaseous disc was studied before the discovery
of extrasolar planetary systems by
\citet{1979ApJ...233..857G,1980ApJ...241..425G} and
\citet{1979MNRAS.186..799L,1986ApJ...307..395L,1986ApJ...309..846L}.
In the linear approximation the planet excites waves at the Lindblad
resonances that deposit angular momentum in the disc.
The flux of angular momentum has different signs in the inner
and outer discs causing the orbital migration of the planet.

\citet{1997Icar..126..261W} proposed that two different types
of planetary drift exist. Type I migration occurs when the planet
mass is small and migrates relative to the disc
with a rate proportional to its mass and the surface density of the disc.
This migration is quite fast and the orbital decay timescale
of the order of $10^5$ years is comparable to the formation timescale of a
giant planet by planetesimal accumulation.
In type II migration the planet is massive enough to open a gap in the
disc. The planet is then locked to the viscous evolution of the disc and
its migration rate will be determined by the strength of the viscosity.
The estimated timescale for type II migration is one or two orders of
magnitude larger than the type I migration timescale for the same
planetary mass. Type II migration is believed to be responsible for
the presence of planets at short orbital distances \citep{2002A&A...394..241T,2003A&A...407..369U}.
Numerical simulations of planet migration in a
viscous disc
\citep[see e.g.][]{2000MNRAS.318...18N,2003ApJ...586..540D}
confirm the inward migration of the planet
on the viscous time-scale
predicted by linear theory
for both accreting and non-accreting planets.

The nonlinear interaction between disc and planet cannot
be fully described analytically or reproduced in laboratory
experiments.
Therefore, multidimensional hydrodynamical simulations of protoplanetary discs
with embedded planets during many orbital periods are necessary to understand
the formation and evolution of extrasolar planetary systems.
However, some differences are found in the simulations
depending on the numerical algorithm employed.
The spiral waves generated around the planet may be 
stationary in the co-rotating frame
\citep[e.g. \zeus{}-based results of][]{1999ApJ...526.1001L}.
Other higher-order hydrodynamical codes show time
variability of the flow in the
spiral arms propagating along the shock.
The quasi-periodic disturbances in the shocks have important
implications for the formation and evolution of vortices 
along the edges of the gap opened by the planet.
In some simulations wavy structures and vortices are observed
at the edge of the gap opened by the planet which interact
with the shocks
\citep[see e.g.][]{2003ApJ...589..556N}.
In this paper, we have used different algorithms presently in use
in the astrophysical community to study the planet-disc system 
in a simple but meaningful case.

%
%
%

%% file: description.tex

\section{Setup description}
\label{sec:setup}

We examined the gap opening by a giant planet in an
infinitesimally thin disk with a constant surface density.
The numerical setup
was defined in the web\footnote{http://www.astro.su.se/groups/planets/comparison/}
where interested modelers were invited to participate in the comparison.

The planet's gravitational potential was given by the formula
\begin{equation}
\phi = \frac {-\mu}{\sqrt{{r}^{2} + \epsilon^{2}}}
\end{equation}
where $r$ is the distance from the planet
and $\epsilon$ is the gravity softening.
The simulations were run with two different softening coefficients:
\begin{equation}
\epsilon_{\text{1}} = 0.2 r_{\text{L}}
\end{equation}
where $r_{\text{L}}=(\mu/3)^{1/3}$ is the size of the Roche Lobe of the planet,
and the larger value
\begin{equation}
\epsilon_{\text{2}} = 0.6 H_{\text{p}},
\end{equation}
with $H_{\text{p}}$ the disc scale height at the planet location.
The second softening was mainly introduced to mimic the
torque cut-off due to the effect of the disc vertical distribution.
The results discussed in this paper concern mostly the calculations that 
use the larger softening.
In our simulations,
the self-gravity, energy transfer and magnetic fields in the
gaseous disc were not considered.

The mass relation between the planet and the star
was chosen so that the reduced mass has the values
$\mu = M_{\text{p}}/(M_{\text{*}}+M_{\text{p}}) = 10^{-3}$ and $10^{-4}$,
corresponding to roughly Jupiter and Neptune masses when
the star mass $M_{\text{*}}=M_{\odot}$.
The planet was kept in a circular orbit
at approximately semi-major axis $a=1$
ignoring the effect of the gravitational torques on the planet.
The position of the planet in the cell
with respect to the cell's corner is given in
Table~1.
The computations were performed in the
radial domain $[0.4a,2.5a]$
to study the influence of the planet
in a sufficiently large fraction of the disc.
In the cylindrical grid codes,
the number of cells
in the radial and azimuthal directions were
$n_{\text{r}} \times n_{\phi}=(128,384)$
with uniform spacing
in both dimensions.
Therefore, the cells around the planet position
were approximately square.
Several tests were done with different schemes at resolution
$n_{\text{r}} \times n_{\phi}=(256,768)$
and $n_{\text{r}} \times n_{\phi}=(512,1536)$
to check the convergence of the results.
The polar coordinates schemes used a corotating reference frame.
The centre of the frame was not specified in the problem description
and codes with frames centred in the centre of mass (CM) and central star
were used.
The star position was fixed at
$r,\phi = (0,0)$ and the planet at
$r,\phi = (1,0)$ in co-rotating coordinates,
where the azimuthal range was $[-\pi,\pi]$.
The Cartesian schemes \adam{} and \pencil{}
were run on a uniform non-rotating grid at resolution
$n_{\text{x}} \times n_{\text{y}} =(320,320)$,
and $n_{\text{x}} \times n_{\text{y}} =(640,640)$.
The computational domain was
$[-2.6a,2.6a]\times[-2.6a,2.6a]$. 
The unit of time used in the simulations was the orbital period
at $a=1$ which is defined as
\begin{equation}
P_{\text{p}}=2 \pi \left[\frac{a^3}{G(M_{\text{*}}+M_{\text{p}})}\right]^{1/2}=2 \pi,
\end{equation}
where $G=1$ and $M_{\text{*}}+M_{\text{p}}=1$.
The angular frequency of the planet was $\Omega_{\text{p}}=1$ in our units.

\subsection{Initial conditions}

The modelled disc was 2-dimensional
so that the vertically integrated quantities 
were solved.
The initial surface density was constant and given by the expression
\begin{equation}
\Sigma_{\text{0}} = 0.002 \frac{M_{*}}{\pi a^2 }
\end{equation}
where $a$ is the semi-major axis of the planet.
We assume that the heat generated by viscous dissipation
and tidal forces in the disc is radiated away, so the disc
remains geometrically thin.
The initial angular velocity
was fixed to the local Keplerian frequency at the given radial position
and the radial velocity was zero initially.

We used the standard sound speed profile of a slightly flaring solar
nebula $H/R = c_{\text{s}}/v_{\text{K}} = 0.05$, 
where $H$ is the disc scale height,
$R$ the distance from the centre of the star,
$v_{\text{K}}$ the local Keplerian velocity,
and $c_{\text{s}}$ the isothermal sound speed defined as 
\begin{equation}
c_{\text{s}}^2=\frac{\partial p}{\partial \Sigma}.
\end{equation}
which has a dependence on radius $c_{\text{s}} \propto r^{-1/2}$.
This corresponds to a locally isothermal equation of state
with a profile $T(r) \propto r^{-1}$
maintained through the simulation.
The disc height at the planet location remains constant during the 
opening of the gap.

The planet mass was gradually increased during the first 5 orbital periods
using the expression
\begin{equation}
\frac{M(t)}{M_{\text{p}}} = \sin^2 \left( \frac{\pi t}{10 P_{\text{p}}} \right)
\end{equation}
to avoid the the appearance of strong shocks seen when the planet
is introduced instantaneously. The gas accretion from the disc
onto the planet was ignored. This situation can be realistic in the case
when the planet's atmosphere fills the Roche lobe and no further
accretion is allowed.

The problem was originally proposed to be run with no artificial
viscosity or as low as possible as allowed by the code.
Some of the used codes
include artificial viscosity to smooth out
the shock fronts and prevent unphysical results
as described in Section~\ref{sec:codedescribe}.
We performed simulations
for each planet mass including a physical viscosity
that generates a stress tensor
with a turbulent viscosity coefficient $\nu$
\citep[see e.g.][]{1959flme.book.....L,1999MNRAS.303..696K}.
The values of the kinematic viscosity used in our simulations
were $\nu=0$ and $10^{-5}$ in units
where $a=1$ and $G(M_{\text{*}}+M_{\text{p}})=1$.
The simulations were run typically
during several hundred orbital periods
for each of the planet masses and viscosity coefficients.

\subsection{Boundary conditions}
\label{sec:boundaries}

To completely define the problem, we describe the
implemented boundary conditions.
The disc was considered as an isolated system with no mass inflow.
We used solid boundary conditions with wave killing zones next
to the boundaries to reduce wave reflection in the cylindrical
coordinates codes.
In the polar coordinates schemes, the damping regions were
implemented in the radial ranges
$[0.4a,0.5a]$ and $[2.1a,2.5a]$, where the following
equation was solved after each timestep:
\begin{equation}
\frac {dx}{dt} = - \frac {x-x_{\text{0}}} {\tau} R(r)
\end{equation}
where $x$ represents the surface density and velocity components,
$\tau$ is the orbital period at the corresponding boundary and
$R(r)$ is a parabolic function which is one at the domain boundary
and zero at the interior boundary of the wave killing zones.
This wave damping condition does not conserve mass, but
the mass loss is very small as shown below.

The grid-based codes in Cartesian
coordinates implemented the same wave killing condition as
the polar codes in the ring $[2.1a,2.5a]$.
Tests were done including the damping condition
in the region $[0.4a,0.5a]$ although this is not necessary
since there is no inner solid boundary.
There was free outflow in the x and y boundaries
in the Cartesian implementations.

Note that the SPH codes implement different
boundary conditions using rings of virtual
particles as described in section~\ref{sec:codedescribe}.

\subsection{Output data}

2-D snapshots of the density and velocity components
were output at 2, 5, 10, 20, 50, 100 and 200
orbital periods for grid codes, although in some cases
the simulations were run up to 500 periods.
All the physical quantities were given at the cell centres.

In the case of SPH codes the output quantities at the previous
times were particle positions, velocity components, smoothing length and mass.
The particles were projected to a 2-dimensional
cylindrical grid with the resolution 
$n_{\text{r}} \times n_{\phi}=(128,384)$
to compare directly with the lower resolution results from the
Eulerian grid codes.
The associated kernel for each particle
used internally by our codes
was the third order spline function introduced by
\citet{1985A&A...149..135M} with a multiplicative
coefficient corresponding to a 2-dimensional simulation.
The density at a given point was calculated by interpolation
with the spline kernel using the expression
\begin{equation}
\langle \rho({\bf r_{\text{i}}}) \rangle= \sum_{j=1}^N m_{\text{j}} W(|{\bf r_{\text{i}}}-{\bf r_{\text{j}}}|,h_{\text{j}})
\label{eq:dens}
\end{equation}
where $m_{\text{j}}$ is the mass of the particle, 
$W(r,h_{\text{j}})$ is the spline kernel and $|{\bf r_{\text{i}}}-{\bf r_{\text{j}}}|$
is the distance from the cell centre to the given particle.
The smoothing length $h_{\text{j}}$ has different
values for each particle.
In a similar manner,
the velocity components were interpolated to the grid 
with the kernel function and normalized with respect to
the integrated kernel.
The resolution element of the SPH models is given by the
smoothing length of the particles.
For the number of particles used in the SPH calculations,
the effective resolution 
is similar to the number of cells in the hydro models
at the aforementioned resolution,
were the particles distributed in an equivalent spatial domain.
\treesph{} uses a smaller
smoothing length than \parasph{} and therefore should have a
slightly better spatial resolution in our calculations.

The azimuthally averaged density was obtained as 
\begin{equation}
\widehat{\Sigma} = \frac{1}{2\pi}\int_{0}^{2\pi} \Sigma\,d\phi
\end{equation}
Slices of the surface density were taken at the planet position
and Lagrangian points in the radial and azimuthal directions.

We calculated the vortensity or potential vorticity,
defined as the ratio of vorticity and surface density
\begin{equation}
\zeta = \frac{(\nabla \times {\bf v})_{\text{z}}}{\Sigma}.
\end{equation}
In the frame rotating with the planet, the vortensity is given by
the expression
$(\nabla \times {\bf v}+2\Omega_{\text{p}})/\Sigma$,
where $\Omega_{\text{p}}$ is the orbital frequency of the planet.

The gaseous disc interacts gravitationally with
the planet by means of the torques
generated by the spiral arms
\citep[see e.g.][]{1979ApJ...233..857G,1984ApJ...285..818P}.
Every few timesteps the contributions from the
inner disc excluding the Hill sphere, outer disc excluding the
Hill sphere and
the torque from the material between 0.5 and 1 Hill radius
to the torque are recorded.
The disc mass interior and exterior to the planet orbit
was also obtained with the same output frequency.

The torques were calculated in units where
$a=1$, $P=2\pi$ and $M_{\text{*}}=1-\mu$
integrating over the corresponding region. 
In the case of a 2-dimensional disc the torque has only a vertical
component which is given by
\begin{equation}
T_{\text{z}} = GM_\mathrm{p} \int{\Sigma\ {\bf r_{p}} \times \frac{\bf r_{\text{e}}}{(r_{\text{e}}^2+\epsilon^2)^{3/2}}} r\,dr\,d\phi
\label{eq:torque}
\end{equation}
where $\Sigma$ is the surface density, ${\bf r_{\text{p}}}$ is the planet position,
${\bf r_{\text{e}}}$ is the distance between the planet and the fluid element.



We performed Fourier analysis of the torque data to understand 
the cause of the observed variability. We used a
Welch windowing function \citep{1992nrfa.book.....P}
to smooth the deviation between the initial and
final amplitudes in the time series.

%% file: codes2.tex
\section{Description of the Codes}
\label{sec:codedescribe}

\begin{sidewaystable*}
\setcounter{table}{1}
\label{tbl:codedescribe}
\medskip
{{\bf Table 1\@.}  Summary of the parameters used in all codes.
Column~2:~name of the users of the code; col.~3:~reference to detailed code description; col.~4:~numerical method (\upwind , \pencillike , \godunov\ or SPH); col.~5:~Courant number used; col.~6:~type of artificial viscosity used (none, von Neumann-Richtmyer, tensor or scalar);
col.~7:~reference frame of hydrodynamic codes (corotating or inertial);
col.~8:~centre of reference frame (centre of mass of star-planet system or primary star);
col.~9:~position of the planet in the cell (centre, corner, arbitrary or coordinates with respect to the cell's corner 
in units of the radial and azimuthal steps $dr$ and $d\phi$);
col.~10:~processor;
col.~11:~approximate execution time on a single processor in hours per 100 orbits.}

\begin{center}
\begin{tabular}{lllllllllll}
\hline
Name & Users & References & Method & Courant & Art. visc. & Frame & Centre & Planet & Processor & $T_{CPU}$ (hr) \\
\hline
\dangelo{} & G. D'Angelo          & 1 &\Upwind\ & 0.5 & none & corot. & CM & corner & Power5 1.65 GHz & 4 \\
\gerben{} & G. Dirksen            & 1 &\Upwind\ & 0.67 & none & corot. & CM & (arb.,0) & P4 2.8 GHz & 6 \\
\cresswell{} & P. Cresswell       & 1 &\Upwind\ & 0.5 & none & corot. & primary & (0.29,0)& P4 2.4 GHz & 11 \\
\kley{} & W. Kley                 & 2 & \Upwind\ & 0.75 & 1.0 (bulk) & corot. & primary & (arb.,0) & P4 3 GHz & 3.4 \\
\fromang{} & S. Fromang           & 3 &  \Upwind\ & 0.5 & 1.0 & corot. & primary & (arb.,0) & Xeon 3 GHz & 16 \\
\fargo{} & F. Masset              & 4,5 & \Upwind\ & 0.5 & 2.0 (vN-R) & corot. & primary & (0.57,0.5) & P4 2.8 GHz & 1.25 \\
\pierens{} & A. Pierens           & 4,5 & \Upwind\ & 0.5 & 1.0 (tensor) & corot. & primary & centre & P4 2.8 GHz & 1 \\
\trampvanleer{} & H. Klahr \& W. Kley & 6 & \Upwind\ & 0.4 & 1.1 (vN-R) & corot. & CM & arb. & Opteron 2 GHz & 16 \\
\pencil{} & W. Lyra & 7 & High-order fin.-diff. & 0.4 & 1.0 (bulk) & inertial & CM & arb. & P4 2.4 GHz & 36 \\
\pci{} & P. Ciecielag \& T. Plewa & 8 & \Godunov\ & 0.8 & none & corot. & primary & (0.57,0) & Xeon 3 GHz & 21 \\
\artur{} & A. Gawryszczak         & 9 & \Godunov\ & 0.8 & none & corot. & primary & (0.57,0.5) & Athlon 2 GHz & 42 \\
\adam{} & A. Peplinski            & 9 & \Godunov\ & 0.7 & none & inertial & CM & arb. & Athlon 1.8 GHz &\\
\trampppm{} & H. Klahr            & 10,11 & \Godunov\ & 0.8 & none & corot. & primary & arb. & Opteron 2 GHz & 28\\
\sijme{} & S-J. Paardekooper \& G. Mellema & 12 & \Godunov\ & 0.8 & none & corot. & primary & arb. & Athlon 1.7 GHz & 25\\
\jupiter{} & F. Masset            & 13 & \Godunov\ & 0.7 & none & corot. & primary & (0.57,0.5) & P4 2.8 GHz & 15.3 \\
\treesph{} & K. Rice              & 14 & SPH & none & bulk + shear & - & CM & arb. & Opteron 1.8 GHz & 10 \\
\parasph{} & C. Sch\"afer \& R. Speith & 15 & SPH & none & bulk & - & CM & arb. & Opteron 2 Ghz & 250 \\
\hline
\end{tabular}
\end{center}

\medskip
{{\bf References:} {\bf 1:}~\citet{1997CPC..101..54};
{\bf 2:}~\citet{1989A&A...208...98K};
{\bf 3:}~\citet{1995CoPhC..89..127};
{\bf 4:}~\citet{2000A&AS..141..165M};
{\bf 5:}~\citet{2000ASPC..219...75M};
{\bf 6:}~\citet{1999ApJ...514..325K};
{\bf 7:}~\citet{2002CoPhC.147..471B};
{\bf 8:}~\citet{2001CPC..138..101};
{\bf 9:}~\citet{2000ApJS..131..273F};
{\bf 10:}~\citet{1993ApJS...88..589B};
{\bf 11:}~\citet{1984JCP...54..174};
{\bf 12:}~\citet{2006A&A...450.1203P};
{\bf 13:}~\citet{1995JCP..120..278};
{\bf 14:}~\citet{1990nmns.work..269B};
{\bf 15:}~\citet{2004A&A...418..325S}.
}
\end{sidewaystable*}

We will now discuss the codes used in the comparison.
Even within the restricted field of astrophysical fluids, there are many different algorithms for computing flows.
There are then different implementations of the same algorithm.
We will therefore start with a discussion of the general principles of various types of codes presented in this paper, and then go on to detail particulars of each code used.
This is not meant to be a general review of all the types of codes used to conduct numerical experiments in astrophysics.
For more detailed information, the reader should refer to any of the plethora of books on the subject \citep[e.g.][]{LaneyCompGasdynam,Toro:1999:RSN,LevequeFiniteVolume}.

The parameters of each code are given in
Table~1, including references in which the algorithms
are described in detail.
Table~\ref{tbl:ResultsLayout}
shows which codes were run for the low resolution defined tests.
In Tables~\ref{tbl:ResultsLayoutHR} and~\ref{tbl:ResultsLayoutHR2}
we show the schemes that were run at higher resolution.

\begin{table}
\centering
\caption{Codes that were run for the lower resolution runs of the setups defined in
Section~\ref{sec:setup}. The grid size for the Eulerian codes was 
$n_{\text{r}} \times n_{\phi}=(128,384)$ for the cylindrical
grid codes and
$n_{\text{x}} \times n_{\text{y}} =(320,320)$ for \adam{} and \pencil{}.
The number of particles was 250\,000 in \treesph{}
and 300\,000 in \parasph{}.}
\begin{tabular}{lcccc}
\hline
& Jupiter & Jupiter & Neptune & Neptune \\
Codes & inviscid & viscous & inviscid & viscous \\
\hline
\dangelo{}      & $\times$ & $\times$ & $\times$ & $\times$ \\
\gerben{}       & $\times$ & $\times$ & $\times$ & $\times$ \\
\cresswell{}    & $\times$ & $\times$ & $\times$ & $\times$ \\
\kley{}         & $\times$ & $\times$ & $\times$ & $\times$ \\ 
\fromang{}      & $\times$ & $\times$ & $\times$ & $\times$ \\
\fargo{}        & $\times$ & $\times$ & $\times$ & $\times$ \\
\pierens{}      & $\times$ & $\times$ & $\times$ & $\times$ \\ 
\trampvanleer{} & $\times$ &   & $\times$ &   \\
\pencil{}       &   & $\times$ &   & $\times$ \\
\pci{}          & $\times$ & $\times$ & $\times$ & $\times$ \\
\artur{}        & $\times$ & $\times$ & $\times$ & $\times$ \\
\adam{}         & $\times$ & $\times$ & $\times$ & $\times$ \\
\trampppm{}     & $\times$ &   & $\times$ &   \\
\sijme{}        & $\times$ & $\times$ & $\times$ & $\times$ \\
\jupiter{}      & $\times$ & $\times$ & $\times$ & $\times$ \\
\treesph{}      &   & $\times$ &   & $\times$ \\
\parasph{}      & $\times$ & $\times$ & $\times$ & $\times$ \\
\hline
\end{tabular}
\label{tbl:ResultsLayout}
\end{table}

\begin{table}
\centering
\caption{Codes that were run at resolution
$n_{\text{r}} \times n_{\phi}=(256,768)$
and equivalent resolutions for the Cartesian
grid and SPH schemes.}
\begin{tabular}{lcccc}
\hline
& Jupiter & Jupiter & Neptune & Neptune \\
Codes & inviscid & viscous & inviscid & viscous \\
\hline
\gerben{}       & & $\times$ & & \\
\cresswell{}    & $\times$ & $\times$ & $\times$ & \\
\pci{}          & $\times$ & $\times$ & & \\
\adam{}         & $\times$ & $\times$ & $\times$ & $\times$ \\
\parasph{}      & & $\times$ & & \\
\hline
\end{tabular}
\label{tbl:ResultsLayoutHR}
\end{table}

\begin{table}
\centering
\caption{Codes that were run
at resolution
$n_{\text{r}} \times n_{\phi}=(512,1536)$.}
\begin{tabular}{lcccc}
\hline
& Jupiter & Jupiter & Neptune & Neptune \\
Codes & inviscid & viscous & inviscid & viscous \\
\hline
\gerben{}       & & $\times$ & & \\
\kley{}       & & $\times$ & & \\
\fargo{}         & $\times$ & $\times$ & $\times$ & $\times$ \\
\hline
\end{tabular}
\label{tbl:ResultsLayoutHR2}
\end{table}


\subsection{Grid Based Codes}
\label{sec:gridcodes}

As the name implies, grid based codes cover the computational volume with a set of grid points at which the various flow variables (velocity, density etc.) are computed.
The mesh geometry (conventionally orthogonal, although this is not absolutely required) can be chosen to reflect the underlying symmetry of the problem.
This often leads to a reduction in the number of grid cells required for a particular problem, and a corresponding cut in computational time.
The codes used in our problem use a reference frame centred in
the centre of mass or primary as indicated in column 8 of Table~1.
All the simulations centred on the primary include the indirect terms
in the potential.
For astrophysical (compressible flow at high Reynolds number) flows, two different approaches to solving the fluid equations are generally used.
However, before we describe these, some general points should be noted.

The most important of these is the Courant-Friedrichs-Lewy (CFL) condition.
Simply stated, information must not travel more than one grid cell per timestep (see, e.g. \citet{1992nrfa.book.....P} for a mathematical derivation).
In a hydrodynamics code, this translates into a restriction on the timestep, based on velocity and sound speed (some authors, e.g. \citet{2004MNRAS.349..678E}, have also added an acceleration condition when appropriate).
Violation of the CFL condition leads to unphysical effects, as causality is violated.
When we refer to the ``Courant number'' in the descriptions below, we are describing an extra safety factor, beyond the formal CFL condition itself.
Note, however, that the CFL condition only applies to time-\emph{explicit} codes.
Implicit solvers are not restricted by it, but no results based on such a code were submitted to us.

Next is the extension to multi-dimensions.
Most algorithms for solving the equations of hydrodynamics have been developed for one dimensional flow.
The conventional method for using a one dimensional algorithm in multiple dimensions is Strang splitting \citep{Strang:1968:CCD}:
solve the 1D equations along each row of cells (the $x_1$ direction), then solve along each column (the $x_2$ direction), using the updated values from the $x_1$ sweep.
Formally, the $x_1$ step should be split in two as $\frac{1}{2}x_1 \rightarrow x_2 \rightarrow \frac{1}{2}x_1$, but most codes do a full step in each direction and alternate which is done first (this is sometimes called ``Godunov splitting'').
The Strang approach makes orthogonal co-ordinates highly desirable.
To minimise the truncation errors this approach produces,
the grid cells must be kept locally square.

Most codes presented here use a rotating polar grid.
For these, there is an extra subtlety: the treatment of the Coriolis force.
As is conventional in fluid dynamics, the simple and obvious way to include this (as an extra force) leads to incorrect angular momentum transport.
Instead, the angular momentum approach of \citet{1998A&A...338L..37K} must be used.
On reflection, this is unsurprising: the Coriolis force simply enforces the conservation of angular momentum in a rotating frame.

Although not relevant to the comparison problem itself, many of the codes here can make use of refined meshes.
High resolution is always desirable, but computationally expensive.
To concentrate grid cells where they are needed, patches of the grid may be calculated at higher resolution, and the results communicated back to the coarser parent grid.
Patches can themselves be patched, giving the potential for extremely high resolution.
If the patches are determined at the start of a calculation, such a code is said to be of the `nested grid' type.
However, some codes can dynamically add and remove patches.
This is known as adaptive mesh refinement (AMR).
For this comparison, we have chosen not to use refined meshes.
This is in the interests of simplicity, since there are a variety of algorithms for performing the refinement, and we are already comparing a large number of codes.
However, we would encourage other workers in the field to compare refinement methods.


\subsubsection{\UPWIND\ Methods}
\label{sec:upwind}

The \upwind\ codes used in this comparison
work by discretising the appropriate version of
the Navier-Stokes equations, and solving that.
These codes use the technique of operator splitting, and some operators are discretised in a finite difference manner, while others are solved with a finite volume method.
For this reason, codes similar to those we shall now discuss are sometimes refered to as `finite difference/volume' schemes, or even just `finite difference.'
We eschew this epithet, since almost any grid code could be described as `finite difference' at some level.

In a typical operator split scheme, each timestep is split into two phases.
During the \emph{source} step, the velocity is updated using the source terms in the Navier-Stokes equations (pressure gradients, gravity etc.).
In the \emph{transport} step, these velocities are then used to advect (the $\mathbf{v}\cdot \nabla \mathbf{v}$ terms) the other quantities.
This is usually done conservatively using the integral form of the equations (integrated over a volume - hence the name).
During the advection step, second order `upwinding' is used (interpolation
based on velocities), to ensure that shocks remain sharp.
Some sort of artificial viscosity is generally required to stop post-shock oscillations making the code unstable.

These codes usually use a staggered mesh, to improve the order of their differencing.
Scalar variables (such as density) are stored at zone centres, while vector quantities are stored at the faces (e.g. $v_1$ is stored at the centre of the $x_1$ face).

Codes like these are often described as being `\zeus{}-like' - a reference to the \zeus{} code of \citet{1992ApJS...80..753S}.
Although that paper provides an excellent description of the methods used, the epithet `\zeus{}-like' does not generally mean ``derived from \zeus{}.''
Rather, they are based on the same or similar algorithms, and \zeus{} happens to be the best known implementation of these.


\paragraph{The \nirvana{} code}

In this comparison, three sets of results were submitted which made use of the \nirvana{} code of \citet{1997CPC..101..54}.
All of the following codes are based on the original version of \nirvana{}, which was not publically released.
Each of the codes was enhanced from the original code base by different groups over a number of years.
Hence, variations between the \nirvana{} codes highlight how even the same
basic algorithm can vary.
Different Courant numbers were also used - \gerben{} used $2/3$, while \dangelo{} and \cresswell{} used $1/2$.


\paragraph{The \kley{} code}

The \kley{} code is two-dimensional mixed explicit/implicit 2nd order \upwind\ algorithm on a staggered grid.
The advection algorithm is based on the monotonic transport scheme by \citet{1977JCP...23..276}.
The \kley{} code can treat radiation transport in the flux-limited diffusion approximation, and includes the full tensor viscosity with dissipation.
In contrast to some other codes the velocity variables that are evolved in \kley{} are radial $v$ and \emph{angular} velocity $\Omega$.
Both radiation and viscosity can be solved implicitly to avoid possible time-step limitations.
We refer the reader to \citet{1989A&A...208...98K} for a full description of the code.
For the purpose of the present calculations the radiation module was replaced by a locally isothermal equation of state.
The viscosity was solved explicitly.
The formulation of the equations, in particular the treatment of the physical and artificial viscosity in the stress tensor components, has been described with respect to the embedded planet problem in detail by \citet{1999MNRAS.303..696K}.


\paragraph{The \fromang{} code}

The \fromang{} code \citep{1995CoPhC..89..127} is derived from \zeus{} \citep{1992ApJS...80..753S}.
The Courant number was 0.5, and an artificial viscosity co-efficient of 1.0 was required to stabilise wave propagation in the disc.


\paragraph{The \fargo{} code}

\fargo{} is a simple 2D polar mesh code dedicated to disc
planet interactions\footnote{\fargo{} is available at
http://www.maths.qmul.ac.uk/$\sim$masset/fargo/}.
It is based upon a standard, \zeus{}-like hydrodynamic solver, but owes its name to the \textsc{Fargo} algorithm upon which the azimuthal advection is based \citep{2000A&AS..141..165M,2000ASPC..219...75M}.
This algorithm avoids the restrictive timestep typically imposed by the rapidly rotating inner regions of the disc, by permitting each annulus of cells to rotate at its local Keplerian velocity and stitching the results together again at the end of the timestep.
The use of the \textsc{Fargo} algorithm typically lifts the timestep by an order of magnitude, and therefore speeds up the calculation accordingly.
The mesh centre lies at the primary, so indirect terms coming from the planets and the disk are included in the potential calculation.
The Courant number was 0.5, and a second order artificial viscosity of $C_2 = 2$ (cf equations~33 and 34 of \citeauthor{1992ApJS...80..753S}) was used.

The standard boundary conditions prescribed in the test problem were used.
In addition, the dependence of the results on the damping condition was
tested using a slightly different boundary were
a transmitted wave boundary condition was used.
The pitch angle of the wake at the inner and outer boundary was
valuated using the WKB approximation.
The content of the border ring was then copied into the ghost
ring, properly azimuthally shifted by the amount dictated by the pitch angle.
This technique is very efficient at removing any reflected wave and yields
similar results to the standard boundary condition defined in
Section~\ref{sec:boundaries}.


\paragraph{The \pierens{} code}

\pierens{} is a 2D code which solves the fluid equations using a \upwind\ method with a time-explicit, operator-splitting procedure.
The \textsc{Fargo} algorithm (see description above) is applied to avoid the timestep limitation at the inner edge of the disc.
Because of this, the code does not alternate radial and azimuthal integrations.
Artificial viscosity is handled by using a bulk viscosity in the viscous stress tensor \citep{1999MNRAS.303..696K}.


\paragraph{The \trampvanleer{} code}

This is a 3D version of \kley{} (see above) with the same second order van-Leer scheme (similar to that used in the \zeus{} and \nirvana{} codes).
\citet{1999ApJ...514..325K} provide a description.
The fact that the code is intrinsically 3D explains why it performs two times slower than the pure \textsc{2d} version \kley{}\footnote{The remaining factor of two comes from the roughly two times smaller Courant number in \trampvanleer{}}.        
We use a moderate value of 1.1 for the von Neumann-Richtmyer type viscosity.
The implementation works in the corotating frame where the centre of mass is the centre of the coordinate system.
Hence no extra acceleration terms are necessary.


\subsubsection{\PENCILLIKE\ Methods}
\label{sec:pencil}

\paragraph{The \pencil{} code}

Pencil is a non-conservative finite-difference code that uses sixth order centred spatial
derivatives and a third order Runge-Kutta time-stepping scheme, being
primarily designed to deal with compressible turbulent
magnetohydrodynamical flows {\footnote { \pencil{} is available at
http://www.nordita.dk/software/pencil-code/}}. Being high-order, Pencil needs
viscosity and diffusivity terms in order to stabilize the numerical
scheme. For this reason, we could not perform inviscid runs.

The code is intrinsically 3D and Cartesian, structured in a
cache-efficient way. The domain is tiled in the y and z direction for
parallelization, with the original 3D quantities being split into 1D
arrays - {\it pencils} - in the x direction, hence the name of the code.
The equations are solved along these pencils in the x direction, which
leads to the convenient side-effect that auxiliary and derived variables use
very little memory as they are only ever defined on one pencil. By calculating
 an entire timestep in the x-direction along the box, Pencil can achieve a
speed-up of $\sim$60\% on typical Linux architectures.

This is the first time Pencil has been applied to the embedded planet problem.


\subsubsection{\GODUNOV\ Methods}
\label{sec:godunovcodes}

The other scheme for grid-based astrophysical fluid flows in common use is that proposed by \citet{Godunov1959}.
Such schemes make use of the fact that there is an analytic solution to the 1D shock tube problem: the so-called Riemann problem.
Godunov's original scheme treated each cell as piecewise-constant (i.e. variables such as density were assumed to be constant throughout the cell), giving a sharp shock at each interface.
\citet{1984JCP...54..174} improved Godunov's method by using parabolic interpolation, giving the `piecewise parabolic method' (PPM) which is the most common implementation in use today.
Implementations of PPM can be in Eulerian or Lagrangian form.
For the purposes of the interpolation, all values are stored at the cell centres (cf the staggered grids mentioned above).
\Godunov\ codes include the pressure gradient in the basic solver.
Since solving the full Riemann problem is computationally expensive, many codes use an approximate solver.
Furthermore, to deal with strictly isothermal flows, a special isothermal Riemann solver must be written, since the conventional one involves $\gamma - 1$ denominators.

\Godunov\ schemes do not usually require any artificial viscosity to ensure stability (sometimes authors will include a small artificial viscosity to prevent post-shock oscillations, but these oscillations do not usually threaten the stability of the code).
Although this is welcome, it should be noted that most implementations contain other `artificial' parts (such as slope limiters used in the interpolations), and any user of a code must bear these in mind.


\paragraph{The \pci{} code}

\pci{} is an AMR code developed by \citet{2001CPC..138..101}.
For the disk-planet interaction problem we used the {\sc Herakles} solver which is an implementation of the PPM algorithm.
{\sc Herakles} was derived from {\sc Prometheus} \citep{1989Preprint449} and provides all the functionality of its predecessor.
The original Riemann solver for complex equations of state was replaced by a much simpler non-iterative (but still exact) version suited for isothermal flows \citep{1994ApJ...420..197B}.
All problems were computed with Courant number of 0.8.
The planet was placed in the corner of a grid cell, to make the grid layout around it as symmetric as possible.


\paragraph{The \flash{} code}

The  \flash{} code \citep{2000ApJS..131..273F} is an AMR code implementing the PPM algorithm in its Direct Eulerian form.
\footnote{\flash{} is available at
http://www.flash.uchicago.edu/}
The Riemann solver was ported from the \pci{} code.
Two sets of results used \flash{}, and we shall refer to these as \artur{} and \adam{}

The \artur{} code was based on release 2.3 of \flash{}.
We patched the code to work as accurately as possible in polar co-ordinates, particularly enforcing the conservative transport of angular momentum.
The Courant number was 0.8 in the simulations presented here.

Instead of running in polar co-ordinates, the \adam{} version of the code used the original Cartesian formulation of \flash{}.
The grid cells were sized to give the same radial resolution,
although since the grid went to $r=0$, the grid size had to be
larger than in the cylindrical schemes to achieve the same resolution.
The code was run at resolution
$n_{\text{x}} \times n_{\text{y}} =(320,320)$
and $n_{\text{x}} \times n_{\text{y}} =(640,640)$.
The boundaries were open and there was free gas flow inside 0.4a.
The damping condition described in Section~\ref{sec:boundaries}
was applied on the outer boundary ring but not in the inner disc.
The Cartesian grid was fixed in space, and the planet and star were free
to move over it (integrated using a simple Runge-Kutta method).
A Courant number of 0.7 was used in the simulations.


\paragraph{The \sijme{} code}

This code uses the approximate Riemann solver suggested by \citet{1981JCP...43....357}, and extended by \citet{1995A&AS..110..587E} to general non-inertial, curvi-linear coordinate systems.
The limiter function used is ``superbee'', and unlike PPM-type approaches, limits the characteristic variables, rather than the primitive variables.
The code uses an AMR scheme similar to {\sc Paramesh} (used in the \flash{} code).
The source terms are handled through the so-called stationary extrapolation method \citep{1995A&AS..110..587E}, which ensures that physically stationary solutions remain stationary.
The equation of state was strictly isothermal.
A full description can be found in \citet{2006A&A...450.1203P}.


\paragraph{The \jupiter{} code}

The \jupiter{} code is a nested grid Godunov code, that can be used in Cartesian, cylindrical or spherical geometry, either in 1D, 2D or 3D.
The \jupiter{} code uses a `two shock' Riemann solver, which assumes that the two waves leaving the interface are shockwaves \citep{Toro:1999:RSN} (the code can also use a `two rarefaction' solver, or a full iterative one).
The rest of the Riemann solver (the sampling of the Riemann fan) is exact.
Assuming that the two waves are shockwaves is not so bad as it might first appear.
Firstly, some initial Riemann states \emph{do} give rise to two shockwaves.
Secondly, the differences from the full Riemann solution are relatively small, so long as the contrast across the interface is not extreme.
In extra tests (not included here), the differences between a `two shock,' `two rarefaction' and full Riemann solver were found to be slight for our comparison problem.
The predictor step (which provides the left and right states of the Riemann problem at the zone interface) is a linear piecewise characteristic method using the monotonized centred slope limiter, and which uses a slope splitting technique \citep{1995JCP..120..278}.
The full viscous stress tensor is conservatively implemented in the three geometries.
No artificial viscosity was required, and the Courant number was 0.7.


\paragraph{The \trampppm{} code}

\trampppm{} is a Lagrangian remap\footnote{Cell boundaries are allowed to move during the advection step, and the results are then interpolated back onto the fixed grid} PPM \citep{1984JCP...54..115} code.
It is based on the routines provided in the \textsc{VH-1} package, modified for accretion disk simulations \citep{1993ApJS...88..589B}.
The modifications involve adding the conservation of angular momentum and equations to treat the evolution of internal energy.
Here always the full Riemann problem is solved iteratively and we approximate the isothermal case with $\gamma = 1 + 10^{-10}$.
In the current isothermal simulations the PPM code does not use any artificial viscosity.
This implementation works in the corotating frame where the star is the centre of the coordinate system, and hence explicitly incorporates the extra acceleration terms due to the offset from the centre of mass.


\subsection{Particle Based Codes}
\label{sec:particlecodes}

Rather than trying to solve the equations of hydrodynamics on a grid, a second group of codes decompose a fluid into small packets of mass (particles), and then follow their evolution.
The method in most common use today is that of Smoothed Particle Hydrodynamics (SPH), developed independently by \citet{1977AJ.....82.1013L} and \citet{1977MNRAS.181..375G}.
We shall describe the basic characteristics of SPH now.
For a more detailed treatment, the reader should consult \citet{1990nmns.work..269B,1992ARA&A..30..543M} and references therein and thereto.

In SPH, each particle's properties are spread (or smoothed) over a small volume of space contained within a smoothing length, $h$.
For example, smoothing out the particle's mass gives its contribution to the density at each point in space.
The smoothing function (or kernel), $W(r,h)$, is not constant, but increases towards the particle's position (assumed to be $r=0$).
In the limit $h \rightarrow 0$, $W$ becomes a $\delta$-function, and perfect fluid behaviour is obtained (with an infinite number of particles).
A Gaussian would be a possible choice for $W$, but compact kernels (where $W=0$ for $r$ greater than some $r_{\text{max}}$) are preferred for computational simplicity.
Particles within the range of the compact kernel are called the neighbours.
When appropriate to the problem, modern SPH codes will allow each particle to have its own smoothing length, chosen to keep the number of neighbours constant (typically a few tens).
The smoothing length is also used to limit the timestep in a way similar to the CFL condition mentioned above.

The major advantage of SPH is that its particle nature makes it fully Lagrangian: there are no advective terms in the equations of motion.
This makes the codes more straightforward to write and understand.
Since high densities imply that more particles are present,\footnote{Although it is possible to let particle masses vary in SPH, it is not entirely trivial to do so} SPH naturally concentrates resolution in high density regions.
Good use can be made of this in collapse simulations \citep[e.g.][]{2003MNRAS.342..926D}.

However, there are disadvantages too.
Foremost is the matter of viscosity.
SPH requires an artificial viscosity to prevent inter-particle penetration, and this tends to make SPH codes quite dissipative.
Resolution can also be a problem in certain calculations.
For example, in the disc calculations presented here, most of the particles are going to be in the outer portions of the disc, and not doing very much.
Also, the details of the gap are of most interest, and SPH will have fewer particles there.


\subsubsection{The \treesph{} code}

This code owes its name to the tree used to locate particle neighbours.
The calculations presented here used 250\,000 particles for the disc, with the star and planet being point masses.
SPH particles that move to within an accretion radius of either the star or planet are accreted \citep{1995MNRAS.277..362B} but in the case of the planet, once the initial ramp up is complete, we do not allow its mass to increase.
We use the standard SPH viscosity \citep[e.g.][]{1992ARA&A..30..543M}, with $\alpha=0.1$ and $\beta=0.2$, but also can include the Balsara switch \citep{1995JCP..121..357} to reduce the shear component of the artificial viscosity \citep[see also][]{2004MNRAS.351..630L}.
A huge saving in computational time is obtained by using individual particle timesteps \citep{1995MNRAS.277..362B} with the time-steps for each particle limited by the Courant condition, a force condition \citep{1992ARA&A..30..543M} and a Runge-Kutta integrator accuracy condition.


\subsubsection{The \parasph{} code}

ParaSPH is a parallelized (using MPI) smooth particle hydrodynamics code.
It follows the approach of \citet{1994ApJ...431..754F}, solving the Navier-Stokes equation including the entire viscous stress tensor.
In contrast to the usual approach of an artificial viscosity of \citet{monaghan:1983}, we use an artificial bulk viscosity.
This allows for an accurate treatment of the physical shear viscosity and for easy comparison to the grid code results, since a constant kinematic viscosity coefficient can be modeled.
Additionally we use the XSPH device to prevent particles from mutual penetration \citep[see e.g.][]{monaghan:1989}
Variable smoothing lengths keep the number of neighbours at 75.
The time integration is performed using a fourth order Runge-Kutta-Cash-Karp integrator for both the particles and the planet.
The code is described in more detail in \cite{2004A&A...418..325S}.

We do not implement exactly the boundary conditions described in
sect.~\ref{sec:boundaries}.
Instead we add virtual particles to the simulation.
They are assigned all physical relevant quantities, such as density, velocity and so on, but are kept in Keplerian orbit about the star.
By their interactions, the virtual particles prevent the SPH particles from escaping.
For the calculations presented here, we used 300\,000 SPH particles and 50\,000 boundary particles.

%% file: results.tex

\section{Results}
\label{sec:results}

\begin{figure}
\includegraphics{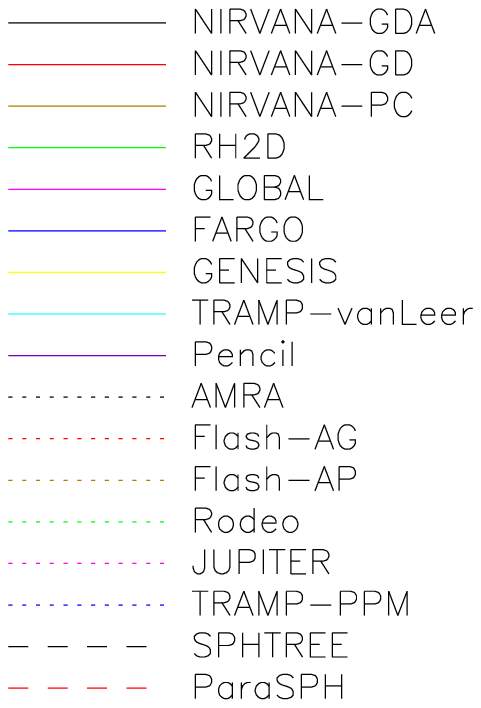}
\caption{Common legend for the comparative plots in
section~\ref{sec:results}.
\Upwind\ codes are represented by solid lines,
\godunov\ codes by dotted lines and SPH codes by dashed lines.}
\label{fig:key}
\end{figure}

In this section we present the results for each of the runs.
The simulations are run for up to 500 orbital periods
using the codes described
in Section~\ref{sec:codedescribe}.
We compare the contours of surface density, vortensity
and averaged density profiles
obtained in the numerical calculations
at several times.
The time evolution of the grid mass and gravitational torque
acting on the planet are shown divided in several contributions.
The Fourier transform of the torques is calculated to
investigate the influence of vortices and disc eccentricity
on the torque acting on the planet.
Several basic properties of the
disc-planet system are discussed
based on the agreement between the codes.
In Section~\ref{sec:highres} we study how the difference between
the codes change as the numerical resolution increases.

The comparative surface density and vortensity maps
are shown for each scheme
in the order they appear in Section~\ref{sec:codedescribe}.
Note that \trampppm{} and \trampvanleer{} were only run
for the inviscid setups,
while \treesph{} and \pencil{} were run for the viscous cases
(see Table~\ref{tbl:ResultsLayout}).
Figure~\ref{fig:key} shows the legend used in
the surface density profiles, mass and torque evolution plots
in this Section.
Different types of algorithms are plotted with different linestyles.


\subsection{Inviscid Jupiter}

\begin{figure*}
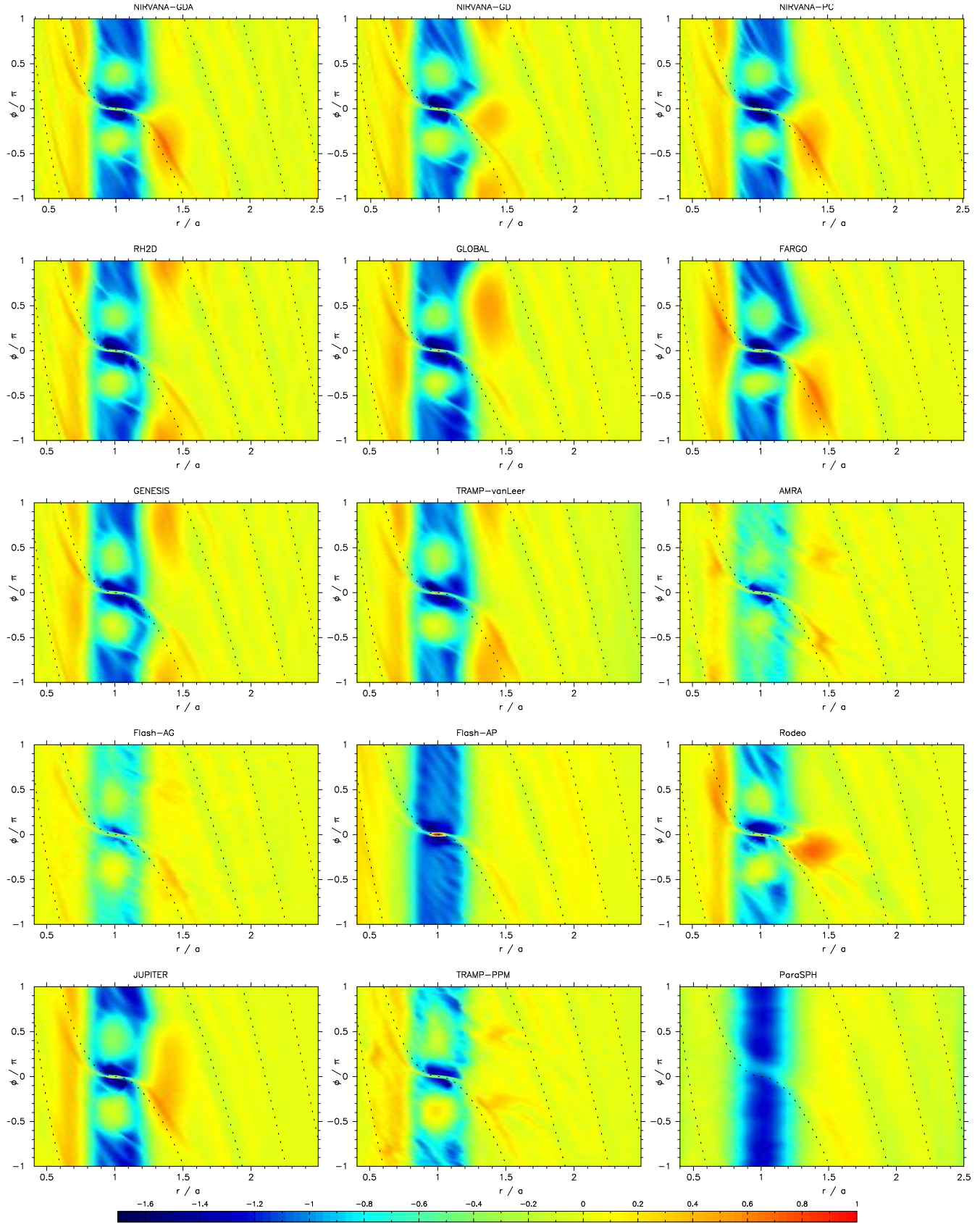

\includegraphics{figs/jup1.dens.100.ps}
\includegraphics{figs/jup.bar.dens.ps}
\caption{Density contours in logarithmic scale
after 100 orbits for the inviscid Jupiter
simulations with overplotted theoretical prediction of
the planetary wake position.
The codes are presented in the same order as
in Section~\ref{sec:codedescribe}.
The \upwind\ methods' results are displayed in the first panels
followed by the \godunov\ codes and lastly the particle based codes.
The density scale ranges between $-1.7 < \log(\Sigma/\Sigma_0) < 1$.}
\label{fig:jupinviscdens}
\end{figure*}

Firstly we consider the case of a Jupiter embedded in an
inviscid disc.
The planet fixed at a given radius opens a deep gap in the disc
as predicted by standard theory
\citep{1986ApJ...307..395L,2000prpl.conf.1135W}.
The contrast between the initial density and the deepest regions in the gap
is about 2 orders of magnitude after 100 orbits.
The planet forms strong trailing spiral arms due to the
differential Keplerian
rotation close to the Lindblad resonances.
Low density regions start to
develop behind the shocks
where the fluid elements encounter the shock at high pitch angle
and change their trajectory.
These regions travel in horseshoe orbits in the corotation
region clearing the gap
as described in \citet{1996ApJS..105..181K}
and creating locked fluid areas at the Lagrangian points
inside the gap.

Figure~\ref{fig:jupinviscdens} shows the density contours
at 100 orbits for all the codes. The dashed line represents
the theoretical position of the shock wave
predicted by \citet{2002MNRAS.330..950O}.
In the Jupiter simulations, the planet mass is too
high for this theoretical estimation
but it allows us to compare the spiral arms pitch angle between the models.
The planetary wakes have a high pitch angle compared
with the theoretical calculation next to the planet.
There is a secondary shock in the Eulerian codes
which starts near the $L_5$ point
and has approximately the same opening angle
as the theoretical prediction.
The secondary shock seems to be related with the density excess inside
the gap behind the planet.
In the outer part of the disc, the pitch angle
of the primary and secondary shocks are very similar.
The existence of secondary shocks and the tightness of the spirals
depend primarily on the equation of state used \citep{1999MNRAS.303..696K}.

There are two density enhancements in all grid-based models
located close to the $L_4$ and $L_5$ points
at azimuthal distance $\Delta \phi = \pm \pi/3$
from the planet.
In the SPH and \adam{} codes the gap is almost completely clean.
Theoretically, the calculation should produce a nearly symmetric
density distribution inside the gap at both sides of
the planet's location for the case of a planet
in a fixed orbit,
which is observed in our results.

\Godunov\ codes that use cylindrical coordinates
such as \artur{}, \pci{} and \trampppm{} have filamentary
structure visible in the disc and the gap possibly due
to the high-order scheme of the codes.
The filaments can be produced by instabilities generated locally
on the corrugated spiral shock.
Their angle does not match the angle of the spiral shocks,
so they cannot be generated around the planet.
In tests performed with the \artur{} code, the filaments appear
in high resolution calculations with larger amplitude but the
same structure.
The \godunov\ codes using a different algorithm than PPM such as
\jupiter{} and \sijme{} do not present filaments although they
seem to have more structure in the disc than
the \upwind\ methods' results.

\begin{figure*}
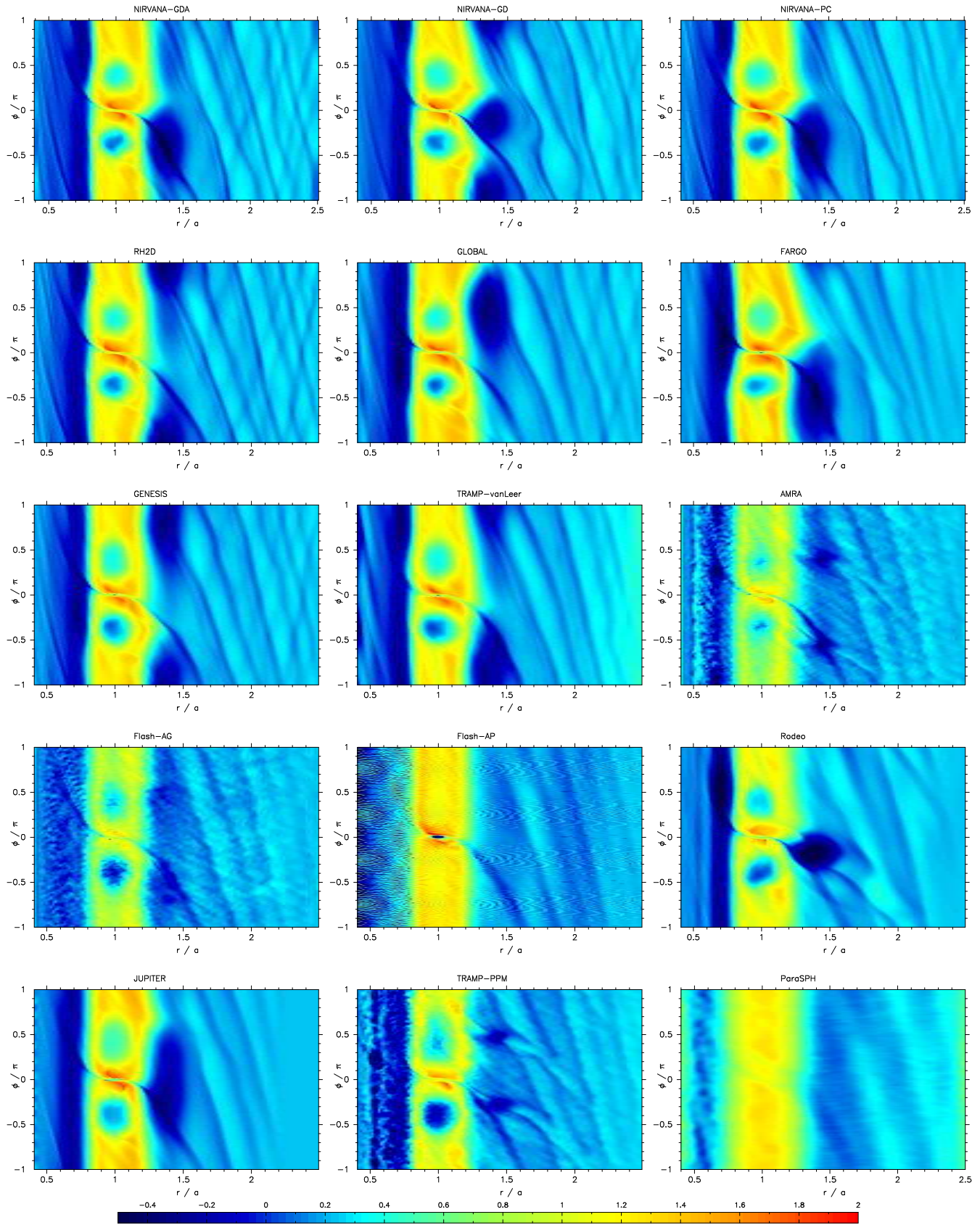

\includegraphics{figs/jup1.vort.100.ps}
\includegraphics{figs/jup.bar.vort.ps}
\caption{Vortensity contours in logarithmic scale
after 100 orbits for the inviscid Jupiter models.
The vortensity range is $-0.5 < \log(\zeta) < 2$.}
\label{fig:jupinviscvort}
\end{figure*}

Figure~\ref{fig:jupinviscvort} shows the vortensity contours calculated
in the corotating frame for the different models.
There are bumps rotating along the edges of the gap opened by the planet
in the grid codes in cylindrical coordinates which survive until 
the end of the simulations.
The resolution does not
permit us to determine whether these density lumps
have locally rotating flow around the core of the vortex.
The vortices are larger in the \upwind\ schemes. 
After 100 periods, most codes show a single bump rotating
along the outer edge,
although \gerben{}, \pci{}, \artur{} and \trampppm{} have two bumps 
which eventually merge by 200 periods.
The knots are dominant in
the \pci{}, \sijme{} and \trampppm{} simulations
and generate their own spiral shocks which extend into the disc.
Most of the codes show one or several smaller density excesses 
at the inner edge.
The vortices in the outer disc 
interact with the planetary shock and generate
quasi-periodic oscillations in the spiral arms.
The oscillations
could also be produced by instabilities near the planet
that interact with the blobs moving along the edge of the gap
and are propagated along the shocks.
Reflected waves appear in the \dangelo{},
\kley{}, \fromang{}
and \pierens{} codes (see Figure~\ref{fig:jupinviscvort}),
despite the use of wave killing boundaries.

In the \parasph{} code the gap edges are less steep than in the case of
the grid based calculations possibly due to the artificial viscosity.
The planetary wake is weak and almost not visible in the inner disc.

\begin{figure}
\includegraphics{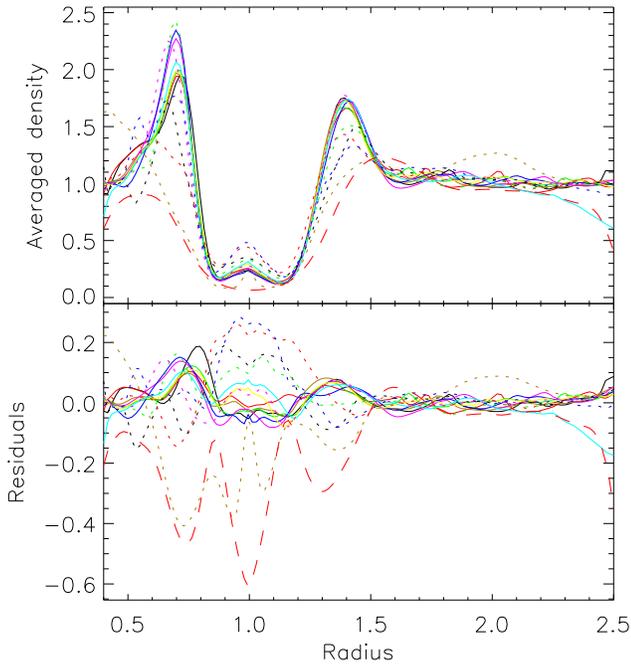}
\caption{The upper panel shows the
normalized surface density profiles averaged azimuthally over $2\pi$
after 100 orbits for the inviscid Jupiter runs.
In the lower panel, the differences between each model and the mean
value are shown in logarithmic scale as
$\log(\Sigma/M_{\text{disc}})-\langle\log(\Sigma/M_{\text{disc}})\rangle$,
where the angle brackets represent the mean. The surface density
has been divided by the disc mass at 100 periods
to remove the dependence on the mass loss due to the boundary
conditions.}
\label{fig:jupinviscprof}
\end{figure}

\begin{figure}
\includegraphics{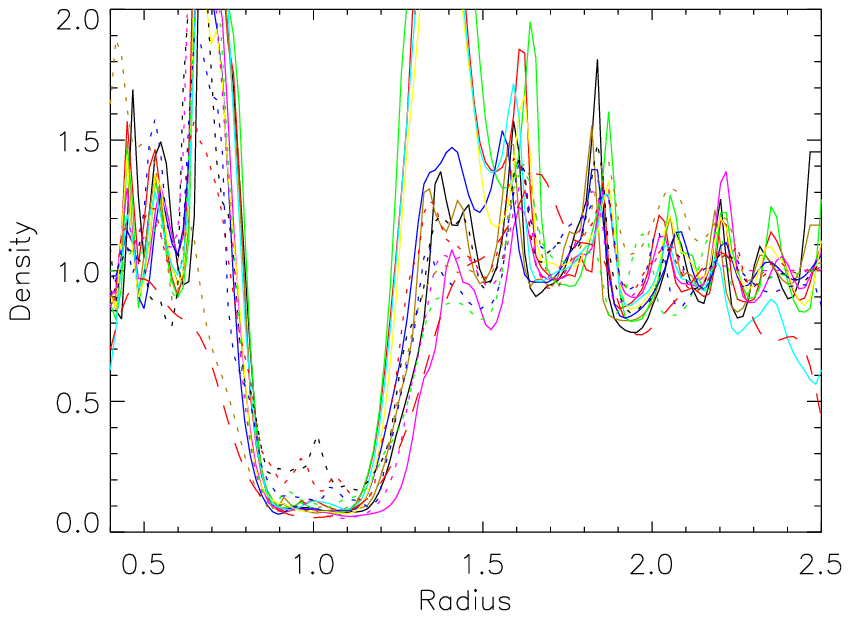}
\caption{Surface density profiles opposite to the planet position 
after 100 orbits for the inviscid Jupiter runs.}
\label{fig:jupinviscprofpi}
\end{figure}
\begin{figure}
\includegraphics{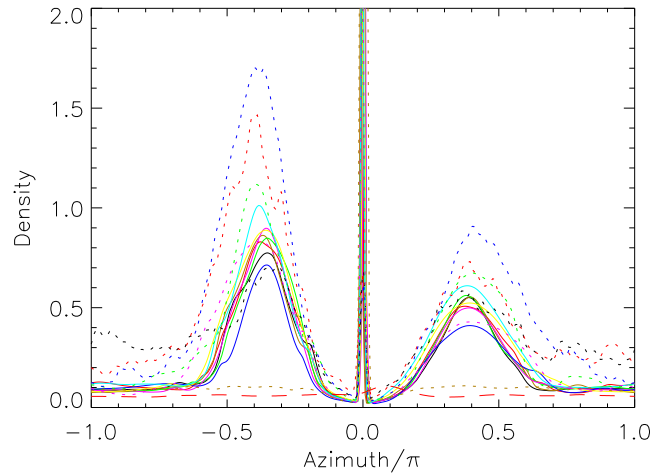}
\caption{Surface density azimuthal slice at the planet radius
after 100 orbits for the inviscid Jupiter calculations.
The trailing Lagrangian point $L_5$ is located at azimuth~$\sim -1/3$
and the leading Lagrangian point $L_4$ is at~$\sim 1/3$ in the
normalized azimuthal units.}
\label{fig:jupinviscphi}
\end{figure}

\begin{figure}
\includegraphics{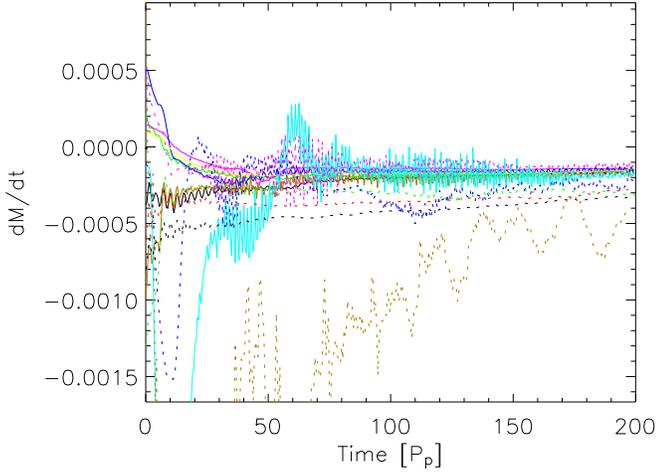}
\caption{Evolution of the disc mass loss rate
over 200 orbital periods for the inviscid Jupiter
simulations.}
\label{fig:jupinviscmass}
\end{figure}

\begin{figure}
\includegraphics{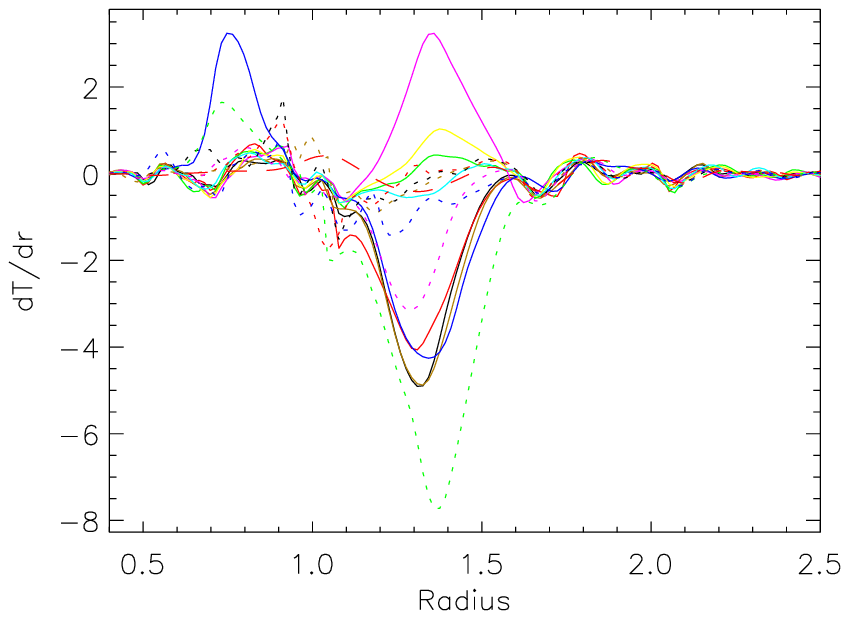}
\caption{Profiles of total specific gravitational torque 
as a function of radius after 100 orbits
for the inviscid Jupiter simulations.}
\label{fig:jupinviscdTdr}
\end{figure}

The azimuthally averaged density profiles and their residuals
normalized by the disc mass
after 100 orbits
are plotted in Figure~\ref{fig:jupinviscprof}.
Note that the relevant portion of the domain is that between
0.5 and 2.1 a, since wave killing conditions are implemented
next to the inner and outer boundaries.  
The depth and width of the gap is in good agreement 
for the grid based models, with a slightly wider gap for
the \pci{} code.
\adam{} has a more depleted inner disc due to the
open inner boundary condition implemented in Cartesian coordinates.
The Cartesian geometry is also more diffusive in this problem and
has a lower resolution close to the
primary compared with the polar coordinates codes.
On the other hand, the shape of the outer disc in \adam{}'s
profiles is similar to the cylindrical codes profiles. 
A wider gap is seen in the \parasph{} simulation.
The oscillations seen in the outer disc are also consistent in
all the codes
with a local maximum in the \adam{} profile at $2a$.
The density peaks close to the edges of the gap --- specially 
in the inner disk --- have a larger spread which
is associated with the size of the vortices.
\Godunov\ codes have smaller vortices in the outer edge
than the \upwind\ methods.
The maximum at the planet location is higher on average for the \godunov\ 
codes.
The \parasph{} code has smoother profiles farther away from the planet
position due to the fact that
the planetary wakes are smeared out.
The residuals of the averaged profiles divided by the disc mass
with respect to the mean
value are shown in the bottom panel
in Figure~\ref{fig:jupinviscprof}.
Since the total disc mass is different after 100 orbits for the various
models, the density profiles normalized by the disc mass have in
general a better
agreement. However, the \parasph{} and \adam{} codes have
most of the mass loss in the inner disc and this method may
artificially increase their residuals in the outer disc.

We plot the density slices opposite to the planet  
after 100 orbits in Figure~\ref{fig:jupinviscprofpi}.
The width and depth of the gap agree well for the different codes
but with a larger dispersion than in the averaged profile.
The amplitude of the peaks at the edges of the gap
differ since there are
vortices which have different sizes and positions
with respect to the planet at a given time.
On the other hand, the size and
the position of the wave crests
agree within a few percent.

In Figure~\ref{fig:jupinviscphi} the azimuthal cuts
of the surface density maps
at the planet radius 
are displayed.
There is a sharp density spike at the planet position.
The shape and prosition of the density bumps at $L_4$ and $L_5$
is slightly asymmetric.
The peak at the trailing Lagrangian point $L_5$ is larger
than at the leading $L_4$ point for all Eulerian codes,
with more conspicuous peaks and larger asymmetry
in the \godunov\ schemes. 
In the \adam{} results there are asymmetric bumps at the Lagrangian points
in the beginning of the simulation, but they have disappeared at 100 periods.
In the \parasph{} calculation the gap is almost completely cleared and
no bumps at the Lagrangian points are observed.
\parasph{} has also a smaller peak at the planet location.


The disc mass loss rate evolution is plotted in Figure~\ref{fig:jupinviscmass}
for the Eulerian codes.
The total disc mass is not conserved due to
the wave damping condition described in section~\ref{sec:boundaries}.
There is a larger mass loss rate in the \adam{} code
owing to the mass accretion in the inner disc but it reaches
an equilibrium value consistent with the cylindrical codes
at late times. 
Some codes gain mass at the beginning of the simulation
and start losing mass after about 10 orbits.
The total mass after 200 orbits is reduced by about
8\% in the \pci{} and \artur{} codes which use \godunov\ algorithms.
Other codes like \dangelo{}, \gerben{} and \kley{}
show a smaller mass loss of about 3\%.
The outer disc mass decreases slowly and in some codes
like \jupiter{} it remains almost constant during 200 orbits.
During the first few orbits, when the gap is not completely cleared,
there is material flowing from the outer to the inner disc perhaps due
to the artificial viscosity.
The inner disc mass shows a strong decrease, especially in the \godunov\ codes.
Despite the spread in mass loss for different codes,
the surface density do not show strong variations between the codes.

The waves excited by the planet deposit angular momentum in the
disc when they are dissipated.
There is an initial smooth phase where the torque increases in absolute
value during the first few orbits
while the planet is growing.
Afterwards, the torques start to display strong oscillations
at the time when the vortices are created.
Vortices grow due to the steep gradients at the gap edges
and through interaction with the planetary wakes. 
We do not have enough time resolution in the density snapshots
to follow the vortex formation and evolution.
The mean value decreases slightly in most codes until the point
when the gap is completely cleared
and stays roughly constant up to 
the end of the simulation.
The effect of the large torque oscillations
on the planet migration
needs to be studied with a free moving planet. 

The torque from within the Roche lobe show significant differences 
between the codes.
The density has a local maximum at the planet
location which depends on the interpolation order of the code,
although the total mass inside the Roche lobe is similar.
The planet is not located in a cell's corner in all codes and
this causes asymmetries in the mass distribution around the planet
In addition, the region close to the planet is not
well resolved at our resolution.
In the following discussion, we compare the torques excluding the
contribution from the Roche lobe.

Figure~\ref{fig:jupinviscdTdr} shows the profiles of
the derivative of the total torque
excluding the Hill sphere with respect to the radius.
The time dependence of the vortices position with respect to the planet
produces a rapidly changing torque. Therefore, the different codes have
different specific torque
profiles at a given time.
The variation appears close to the gap edges where most of the
angular momentum is deposited.
The differences are larger
at the outer edge position where the vortices are bigger
than at the inner edge.
On the other hand, farther away from the planet position
the torques are remarkably similar for all the codes.

\begin{figure*}
\begin{tabular}{ccc}
\includegraphics{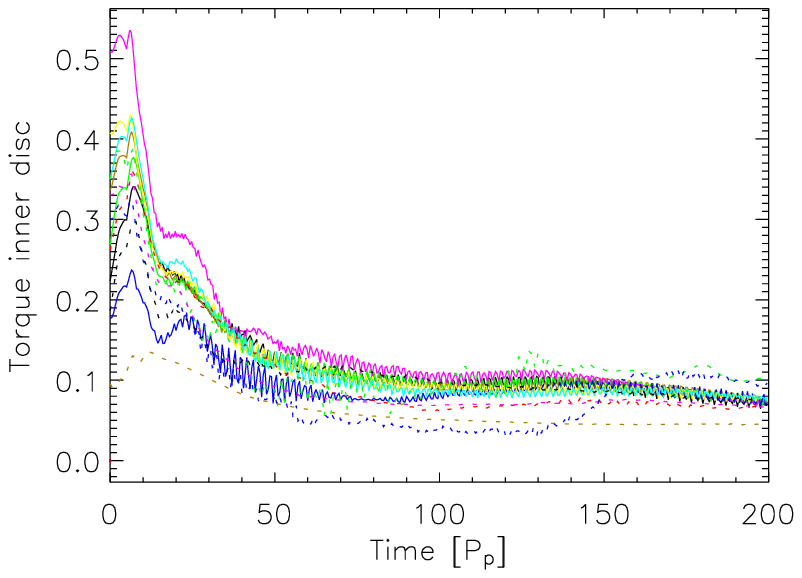} & \includegraphics{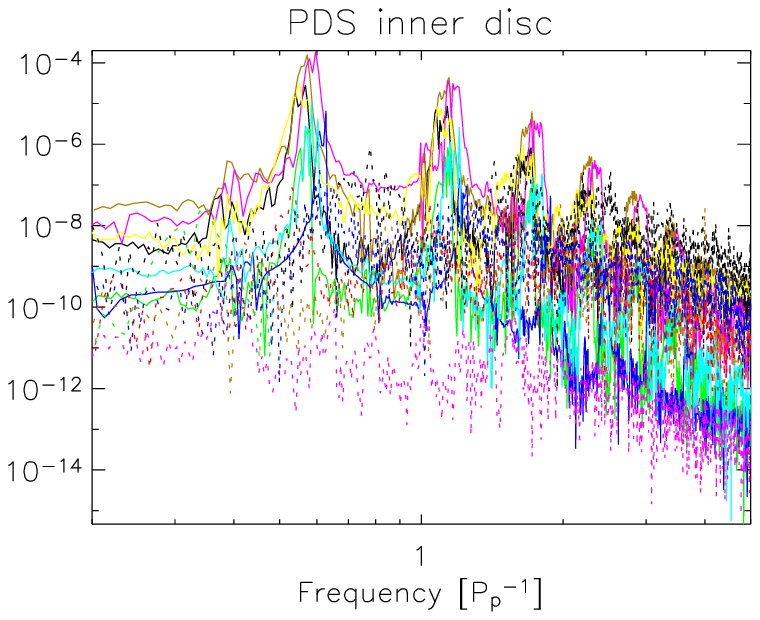} \\
\includegraphics{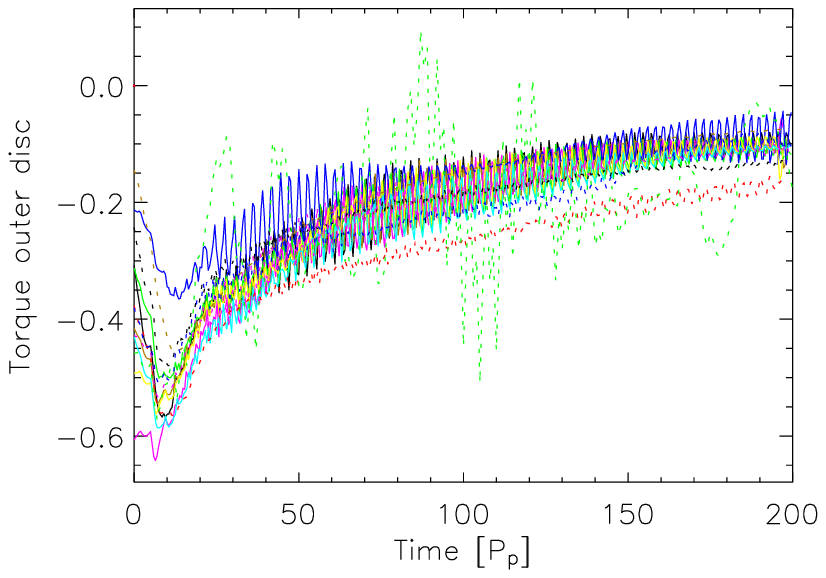} & \includegraphics{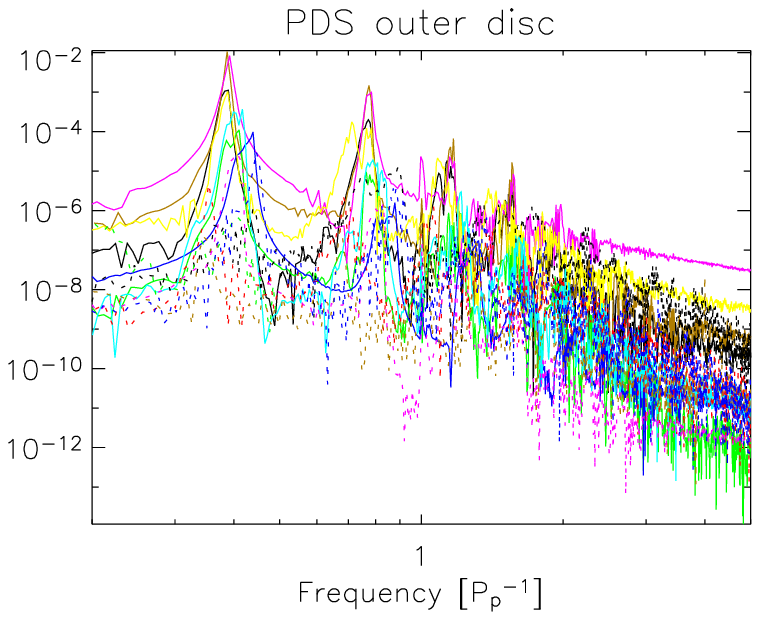} \\
\includegraphics{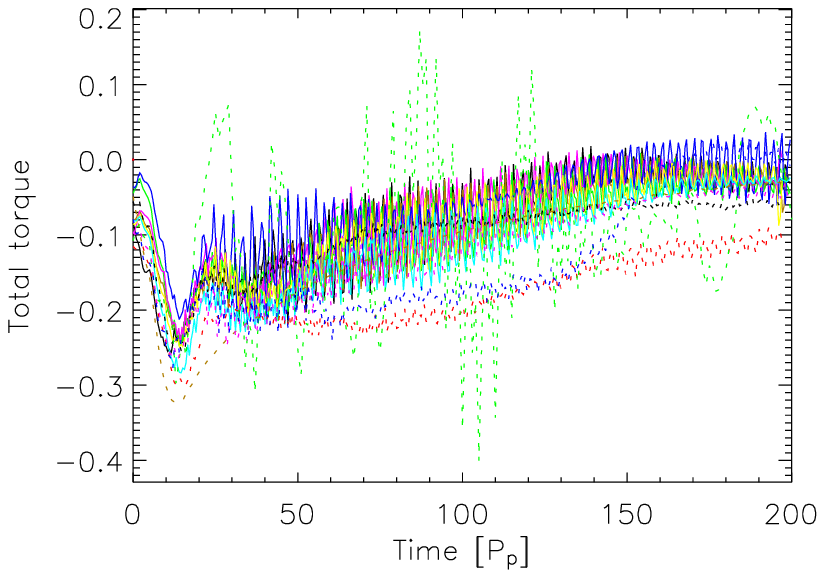} & \includegraphics{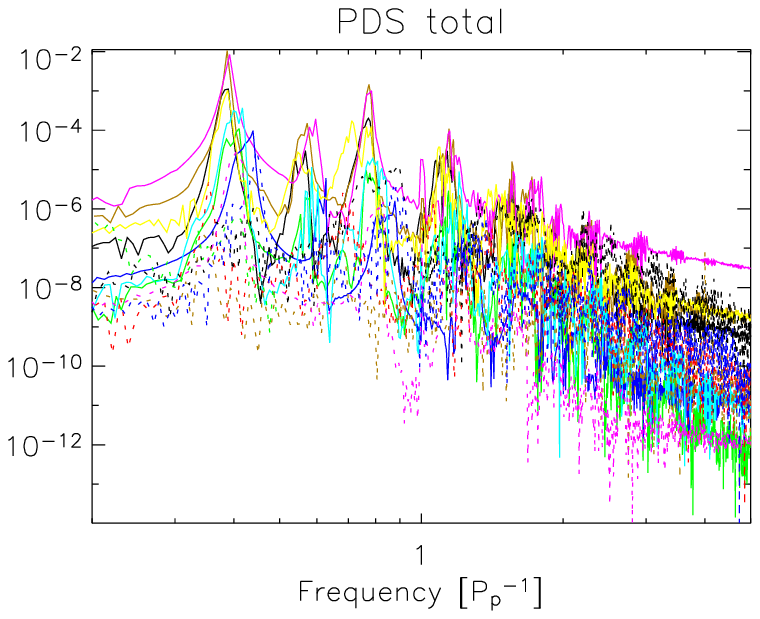}
\end{tabular}
\caption{Torques time series for the inviscid Jupiter simulations
smoothed over 10 periods and the corresponding
normalized power density spectra in logarithmic scale.
The upper panels show the torque contribution from the inner disc,
the middle panels show the torque from the outer disc
and in the bottom panels the total torque is plotted.
In all plots the contribution from inside the
Hill sphere is ignored to avoid numerical noise.}
\label{fig:jupinviscpds}
\end{figure*}

The time evolution of the gravitational torque acting on the planet is
shown in Figure~\ref{fig:jupinviscpds} divided in inner, outer disc
and total contributions.
A running time average
over 10 orbital periods
has been applied to the data to avoid large oscillations.
The torque contribution from the disc material
inside the planet's orbit gives
a positive torque on the planet which tends to drive
the planet outwards in all models, while the torque from the material
outside the planet's orbit pushes the planet towards the star.
The outer disc contribution is dominant
and gives a total negative torque
which takes away angular momentum from the planet
and would cause inwards migration in case the planet
were released.
The torque order of magnitude and sign agrees for all the codes 
after 200 orbits
except for \trampppm{} which has a value close to zero.
The averaged values at the end of the simulation are given
in Table~\ref{tbl:torques.jup1}.

\begin{table}
\centering
\caption{Averaged gravitational torques
between 175--200 periods
in units where
$a=1$, $P=2\pi$ and $M_{\text{*}}=1-\mu$
for the Jupiter inviscid simulations at the
end of the simulations. The time is given
in orbital periods of the planet.}
\begin{tabular}{lc}
\hline
Code & Torque \\
\hline
\input{torques.200.jup1.dat}
\hline
\end{tabular}
\label{tbl:torques.jup1}
\end{table}

The power density spectra (PDSs) of the corresponding gravitational
torque components
are shown on the right hand side plots in
Figure~\ref{fig:jupinviscpds}.
The panels show the low frequency part of the power spectrum
in logarithmic scale.
The semi-periodic oscillations
associated with vortices rotating along
the gap edges are present in the PDSs for models
where blobs appear in the density maps next to the gap.
Several peaks appear in the outer disc PDS with
frequency corresponding to
roughly 0.4 times the planet's orbital frequency
with several harmonics.
This frequency is the difference between the planet's orbital
frequency and the angular velocity of the high density regions
moving next to the gap.
Assuming that the density lumps orbit the central star with
Keplerian speed,
the position of the blob estimated from the PDS frequency
is about $1.4a$, in agreement
with the center of the blobs observed in the density maps.
The harmonics possibly appear because the potential
of an extended density blob is not sinusoidal and
creates amplified multiple frequencies.
In some codes, there are several vortices next to the outer edge
which perturb the planet with the same frequency but different phase.
In the inner disc contribution PDS,
there are several codes with peaks at
about 0.7 times the planet's orbital frequency
and its harmonics.
The estimated blob position is about $0.7a$, which again
agrees with the center of the vortices observed next to
the inner edge of the gap in the density maps.

The PDSs of the torque from the material inside the Roche lobe
show high frequency quasi-periodic
variations at about 20 times the planet orbital frequency.
This high frequency oscillations may be caused by the
circumplanetary disc which makes several orbits around the planet
within the planet orbital period
although the region is poorly resolved at our resolution.
There is a local
maximum in density
inside the Roche lobe
and the material gives a leading contribution
to the total torque acting on the planet.



\subsection{Viscous Jupiter}

\begin{figure*}
\includegraphics{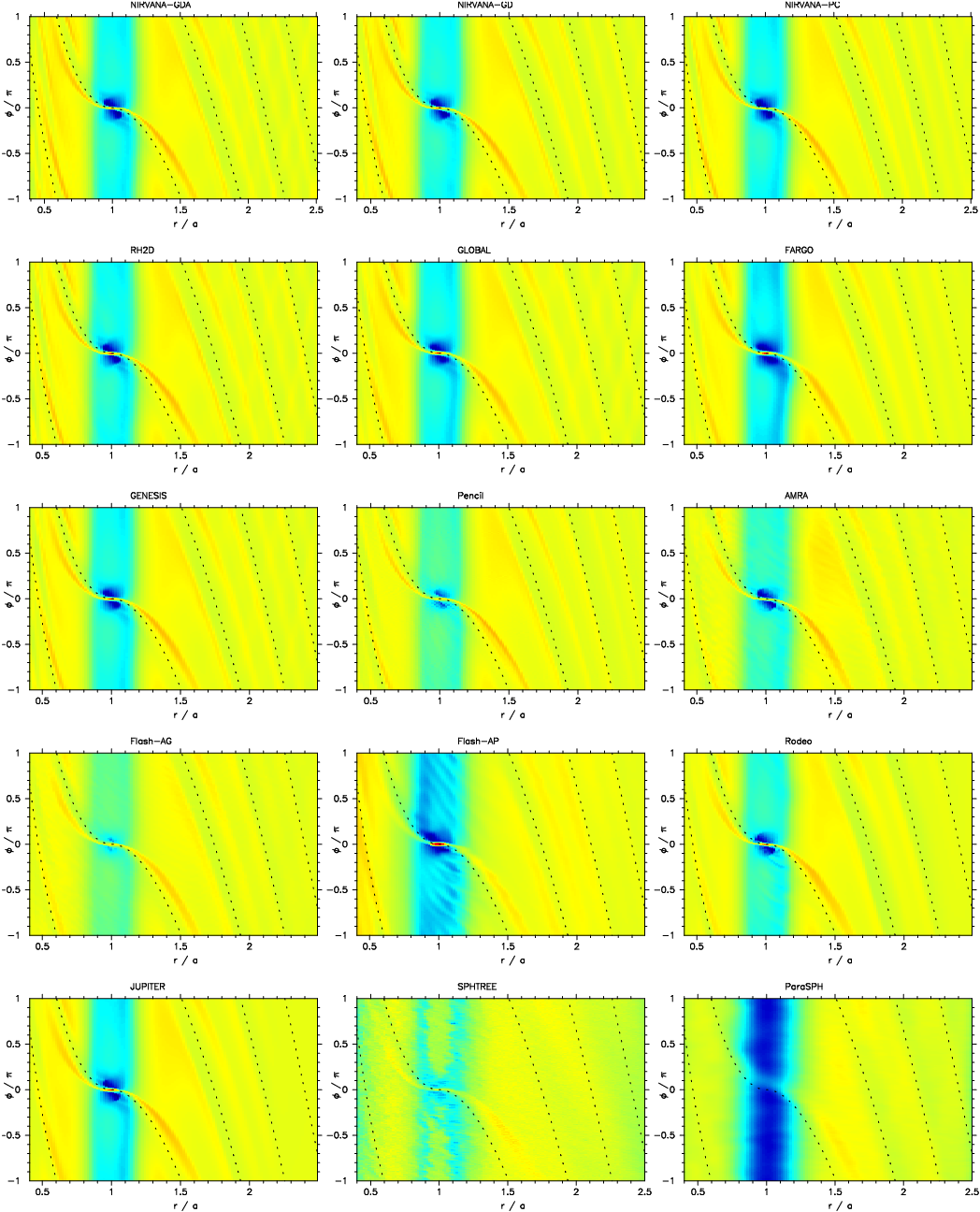}
\includegraphics{figs/jup.bar.dens.ps}
\caption{Density contours after 100 orbits for the viscous Jupiter
simulations. 
The dashed line is the estimated theoretical position of the planetary shocks.
The density range is again $-1.7 < \log(\Sigma/\Sigma_0) < 1$.}
\label{fig:jupviscdens}
\end{figure*}

\begin{figure*}
\includegraphics{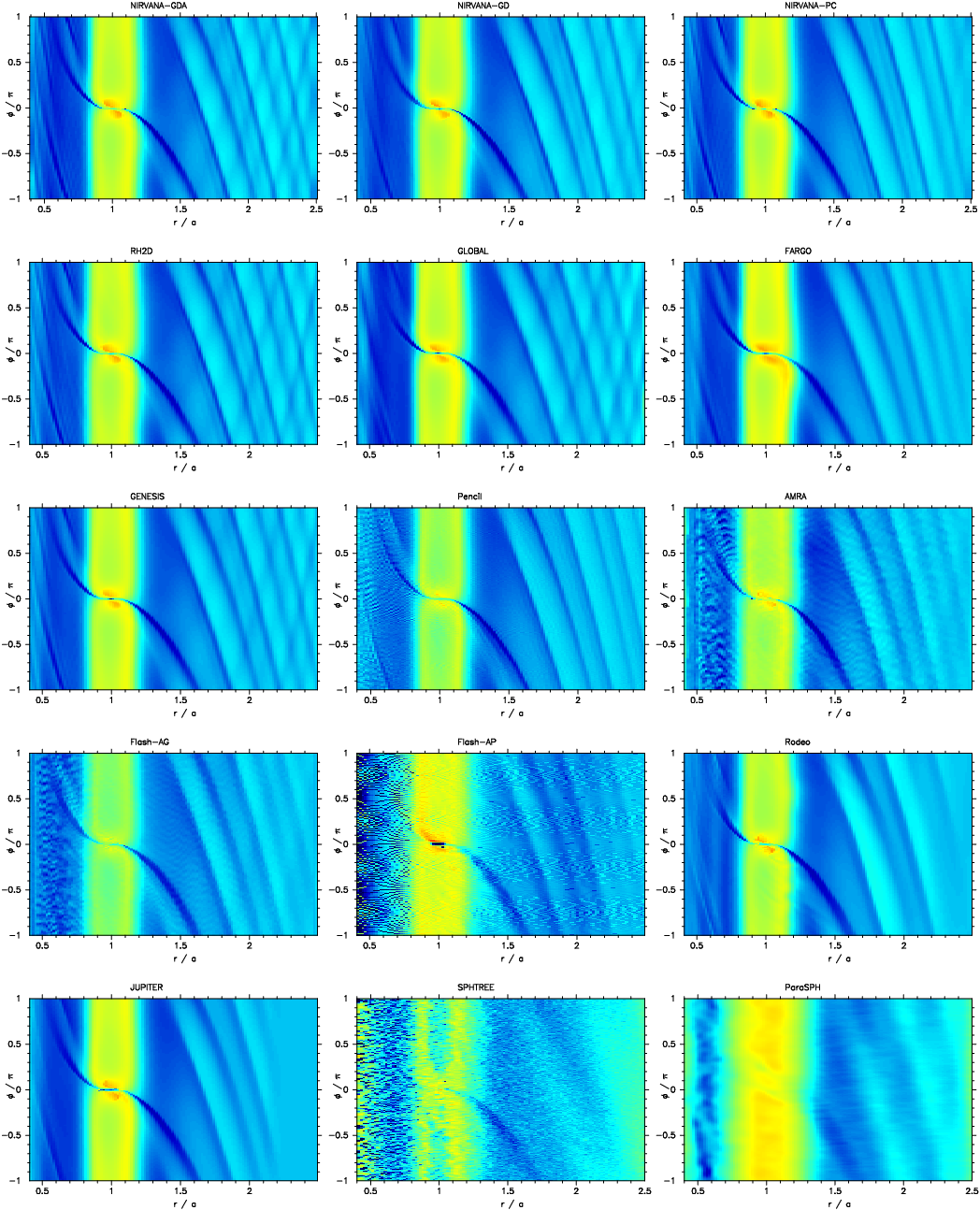}
\includegraphics{figs/jup.bar.vort.ps}
\caption{Vortensity maps for the viscous Jupiter case after 100 orbits.
The logarithmic scale is $-0.5 < \log(\zeta) < 2$.}
\label{fig:jupviscvort}
\end{figure*}

The density contours for Jupiter in a disc with
Navier-Stokes viscosity $\nu=10^{-5}$ 
are shown after 100 orbits for all the codes
in Figure~\ref{fig:jupviscdens}.
The planet opens a narrower gap in the disc 
in this case.
The flow is much smoother than in the inviscid calculation
and the blobs moving along the gap are not observed.
The density enhancements 
seen at the Lagrangian points inside the gap in the inviscid calculations
are not present.
The spiral arms generated by the planet are stationary.
The filamentary structure that appeared
in the inviscid Jupiter runs in the \godunov\ codes
is reduced in amplitude. The reduction is stronger in \artur{} and \adam{}
results than in \pci{} which uses a different dissipation algorithm.

In Figure~\ref{fig:jupviscvort}, we plot the vortensity
for the viscous Jupiter case. The maps are smooth
compared with the inviscid simulations and vortices are
not visible in the logarithmic scale.
Reflected waves are visible in the \dangelo{}, \kley{}, \fromang{}
and \pierens{} results despite the use of the wave killing zones.

\begin{figure}
\includegraphics{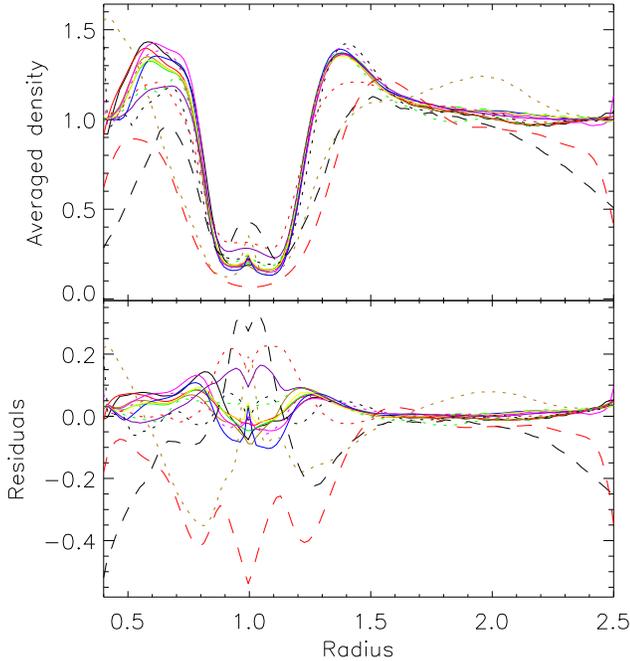}
\caption{The upper panel shows the
surface density profiles averaged azimuthally over
the whole azimuthal range
after 100 orbits for the viscous Jupiter case.
In the lower panel, the difference between each model and the mean
value is shown as defined in
Figure~\ref{fig:jupinviscprof}.}
\label{fig:jupviscprof}
\end{figure}

\begin{figure}
\includegraphics{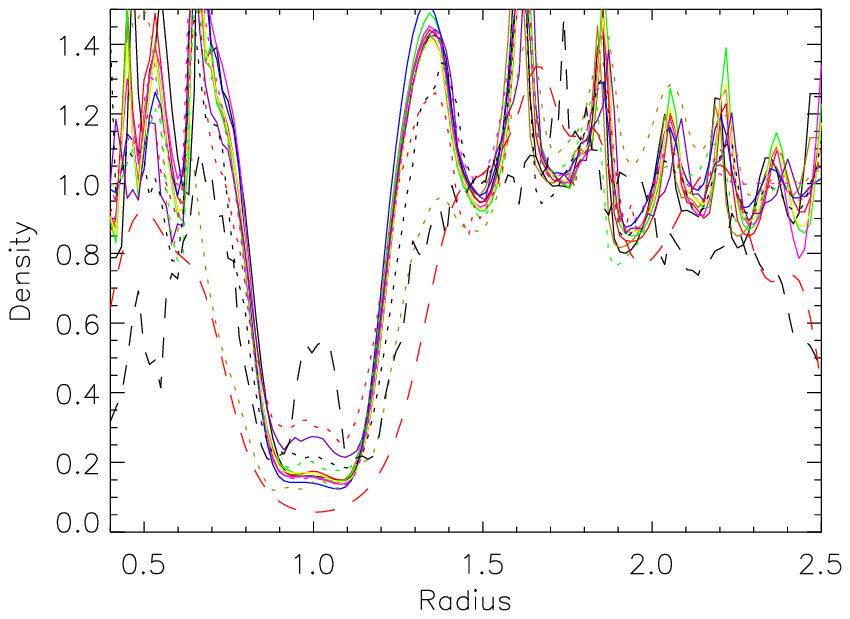}
\caption{Surface density profiles opposite to the planet position 
after 100 orbits for the viscous Jupiter runs.}
\label{fig:jupviscprofpi}
\end{figure}

\begin{figure}
\includegraphics{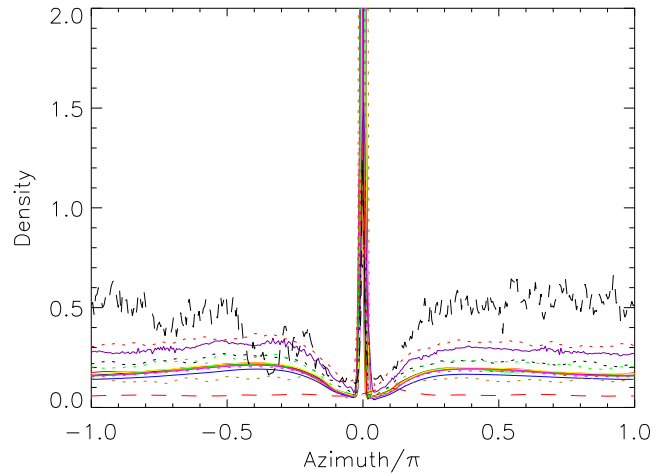}
\caption{Surface density azimuthal cut at the planet position
after 100 orbits for the viscous Jupiter runs.}
\label{fig:jupviscphi}
\end{figure}

\begin{figure}
\includegraphics{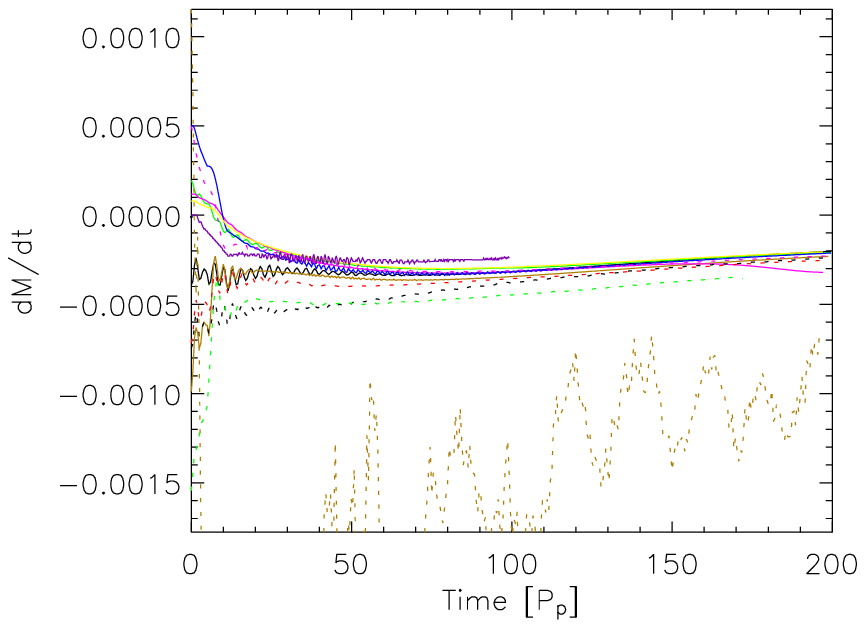}
\caption{Evolution of the
disc mass loss rate 
over 200 orbital periods for the viscous Jupiter case.}
\label{fig:jupviscmass}
\end{figure}

In Figure~\ref{fig:jupviscprof} we show
the azimuthally averaged density profiles 
and normalized residuals after 100 orbits.
The depth and width of the gap agree well
for the grid codes
with a shallower gap in the \adam{} code. 
The gap is wider and deeper in the \parasph{} simulation.
The \treesph{} code has a small peak at the planet radius
and the inner disc is depleted due to mass loss.
An slightly asymmetric gap structure is observed in most codes
with a deeper dent 
outside the planet's orbit.
The oscillations in the outer disc have disappeared 
in the grid codes or have been reduced considerably by the viscosity.
The differences of the averaged profiles
with respect to the mean value
are shown in Figure~\ref{fig:jupviscprof}.

We plot the surface density profiles at $\phi=\pi$
after 100 orbits in Figure~\ref{fig:jupviscprofpi}.
The peaks at the edges of the gap agree well 
since there are no vortices in the viscous runs
and the spiral arms are weaker.
Due to the viscosity, the gap is
narrower and shallower than in the
inviscid case.
The shape of the spiral arms agree within a few percent
for the grid based codes. 
The SPH codes agree in the general shape of the density
profile but have weaker spiral waves.
\treesph{} has a density peak in the middle of the gap
opposite from the planet.

In Figure~\ref{fig:jupviscphi} we plot the azimuthal cuts of the
surface density maps at the centre of the gap. 
A sharp density spike is seen at the planet position
in all codes.
The density bumps at the equilibrium points inside the gap have disappeared.
Most of the grid codes show a constant density of
about 15\% of the initial value.
\parasph{} has a lower density than the grid codes, while \treesph{}  shows
a density
about twice as large as the grid codes.
The presence of oscillations
in the \treesph{} azimuthal profile 
may be explained because the number of particles is too small
to resolve the gap.
The effective resolution after projection in the radial
range $[0.4a,2.5a]$ is smaller than in the grid simulations
since the radial domain extends until $10a$ and
at the end of the simulation a significant fraction
of the particles has been accreted.

We plot the evolution of the disc mass loss rate
in Figure~\ref{fig:jupviscmass}.
There is less mass loss than in the inviscid Jupiter
case due to the weaker waves and
the agreement in the loss rate is generally very good.
\adam{} has a larger mass decrease due to
the open inner boundary.
\pencil{} has a very small mass loss possibly due to the freezing zones
in the boundaries.
The total mass loss after 200 orbits shows better agreement than
in the inviscid case.
\Upwind\ methods show a reduction of about 5\% of the initial mass,
while \sijme{}, \pci{} and \artur{}
codes lose close to 8\% of their mass.
During the first few orbits,
there is again gas flow from the outer to the inner disc
when the gap is not cleared. 
The outer disc mass decreases slightly for some codes while others present
an increase of roughly 1\%.
There is a substantial decrease in the inner disc mass with 
an agreement of approximately 10\%
between the different models.


The amplitude of the torque oscillations
are smaller compared with the inviscid runs.
There is again an initial stage where the torque increases in absolute
value while the planet mass is increasing.
The torques start to oscillate
at about 10 orbits and later
possibly due to the formation of small vortices or eccentricity of the disc.
In most codes the oscillations decrease and become
very small by the end of the simulation.

\begin{figure}
\includegraphics{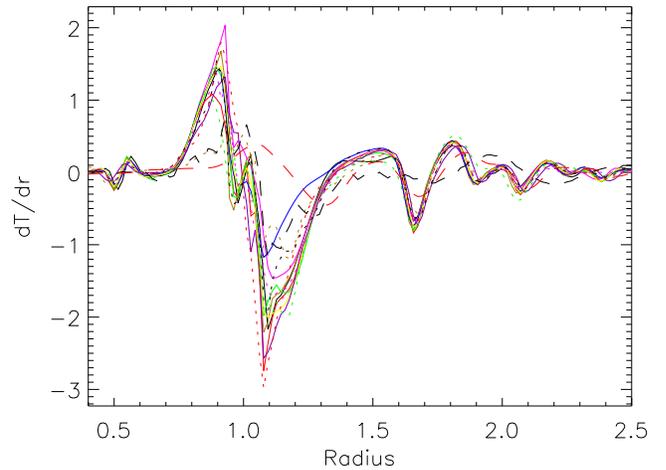}
\caption{Profiles of total specific torque acting on the planet
after 100 orbits
for the viscous Jupiter case.}
\label{fig:jupviscdTdr}
\end{figure}

In Figure~\ref{fig:jupviscdTdr} the profiles of
the specific total torque
excluding the Hill sphere are shown.
The profiles show a much better agreement than in the
inviscid Jupiter simulations. In the viscous case,
vortices are not observed in the density maps after 100 orbits
and the torque radial profiles are not time dependent.
There is a dominant contribution from the corotating region
in the grid-based schemes from the exchange of angular
momentum with gas flowing in horseshoe orbits,
although the material inside the Roche lobe is
not considered.
The outer disc gives a negative torque contribution on the planet
driving inwards migration
and the inner disc produces a positive torque that pushes
the planet outwards.
The profiles of the polar coordinates codes agree
within a few percent.

\begin{figure*}
\begin{tabular}{ccc}
\includegraphics{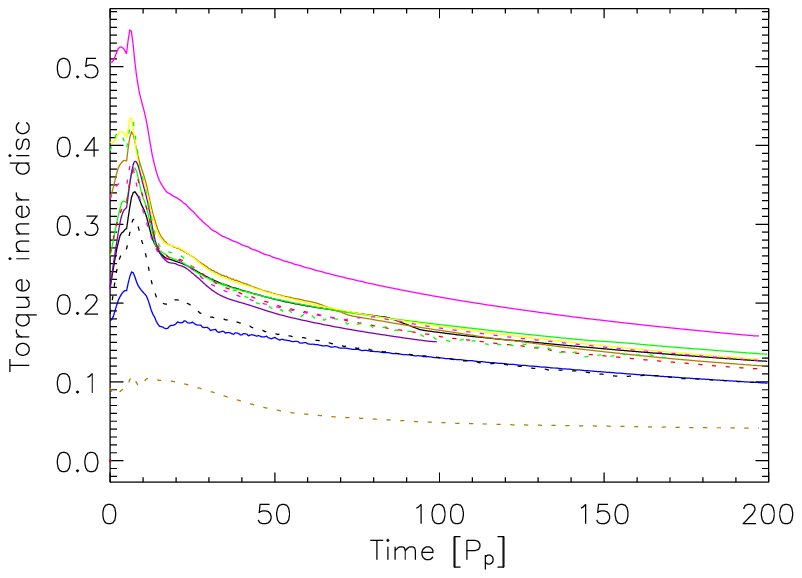} & \includegraphics{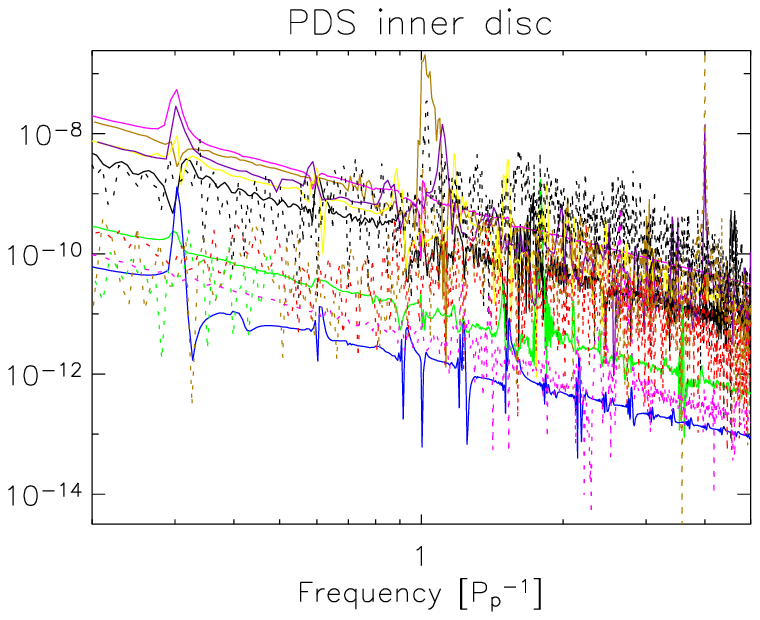} \\
\includegraphics{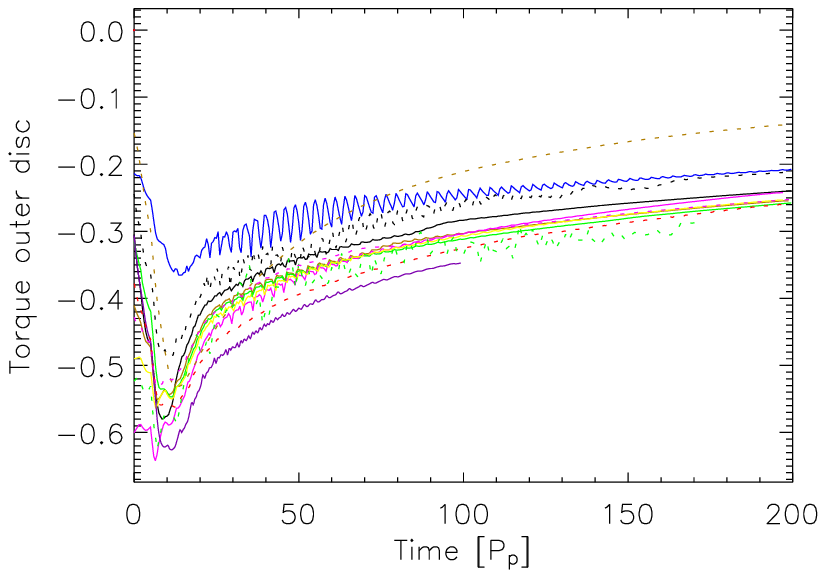} & \includegraphics{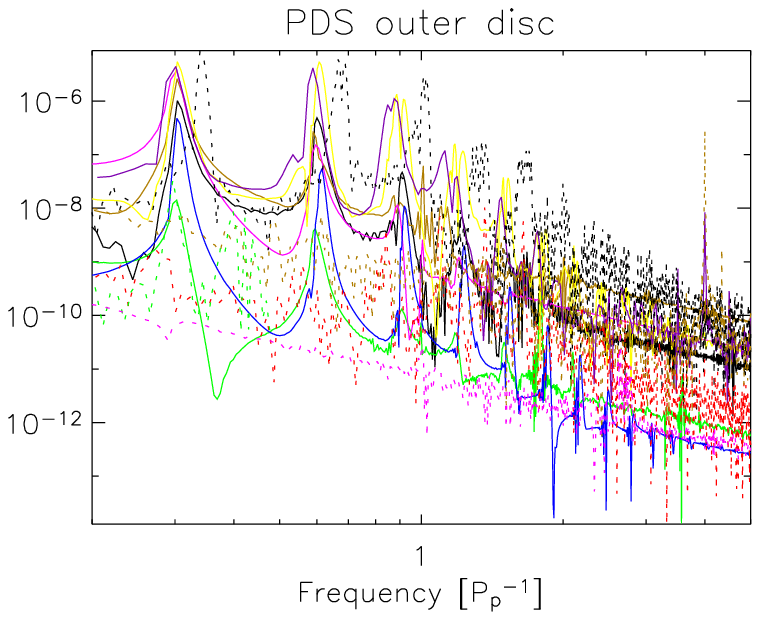} \\
\includegraphics{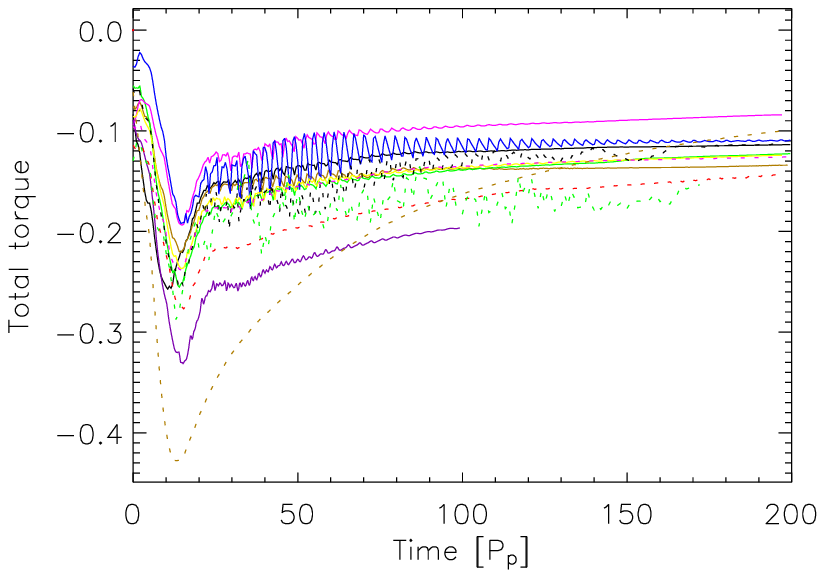} & \includegraphics{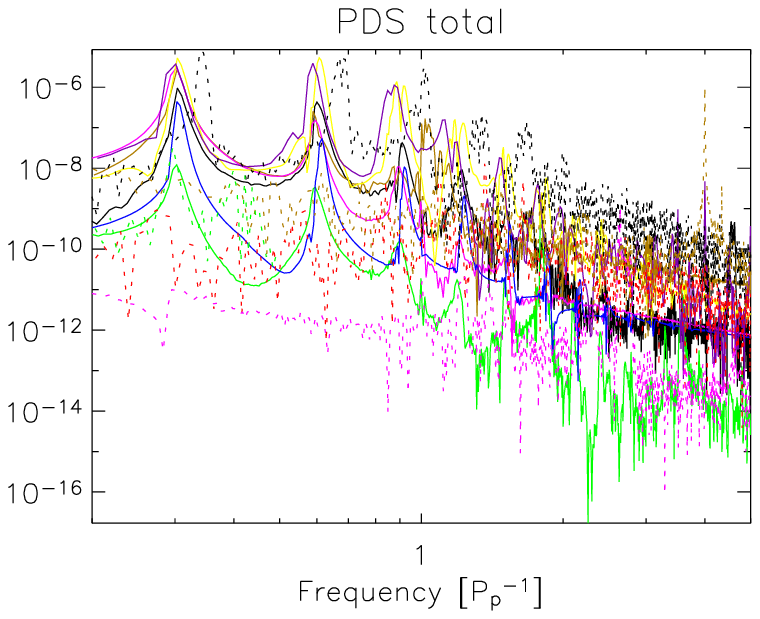}
\end{tabular}
\caption{Running time averaged torques for the viscous Jupiter simulations
and the corresponding PDSs of the raw data. 
The plots are shown in the same order as in Figure~\ref{fig:jupinviscpds}.
All the figures exclude the Roche lobe contribution.}
\label{fig:jupviscpds}
\end{figure*}

The time average of the torque acting on the planet and
their PDSs are 
shown in Figure~\ref{fig:jupviscpds}.
The outer disc torque contribution is again dominant
and gives a negative total torque.
The total averaged torques at the end of the simulation
are shown in Table~\ref{tbl:torques.jup2}.
The PDSs of the different torque contribution 
are shown in the right hand side panels in Figure~\ref{fig:jupviscpds}.
The plots show the low frequency part of the PDSs 
in logarithmic scale.
There is a peak at 0.3 times the planet's orbital frequency
and several multiples in the outer disc PDS.
In some models, there is also a small peak at the same
frequency in the PDS form the inner disc.
This quasi-periodic oscillations may be produced by vortices
appearing during the first orbits of the simulation
and eventually removed by the viscosity.
Other possible explanations are
asymmetry in the edge of the gap
or slight eccentricity of the disc.

The torque from the gas inside the Hill sphere 
presents again a power spectrum with high frequency peaks
at several times the Keplerian frequency at the
planet radius.
The smoothing length is close to half of
the Hill radius and
the resolution in the Roche lobe is low
to study the possible
presence of a circumplanetary disc
rotating at high angular frequency.


\begin{table}
\centering
\caption{Averaged torques between 175--200 periods
in units where
$a=1$, $P=2\pi$ and $M_{\text{*}}=1-\mu$
for the Jupiter viscous simulations.}
\begin{tabular}{lc}
\hline
Code & Torque \\
\hline
\input{torques.200.jup2.dat}
\hline
\end{tabular}
\label{tbl:torques.jup2}
\end{table}


\subsection{Inviscid Neptune}

\begin{figure*}
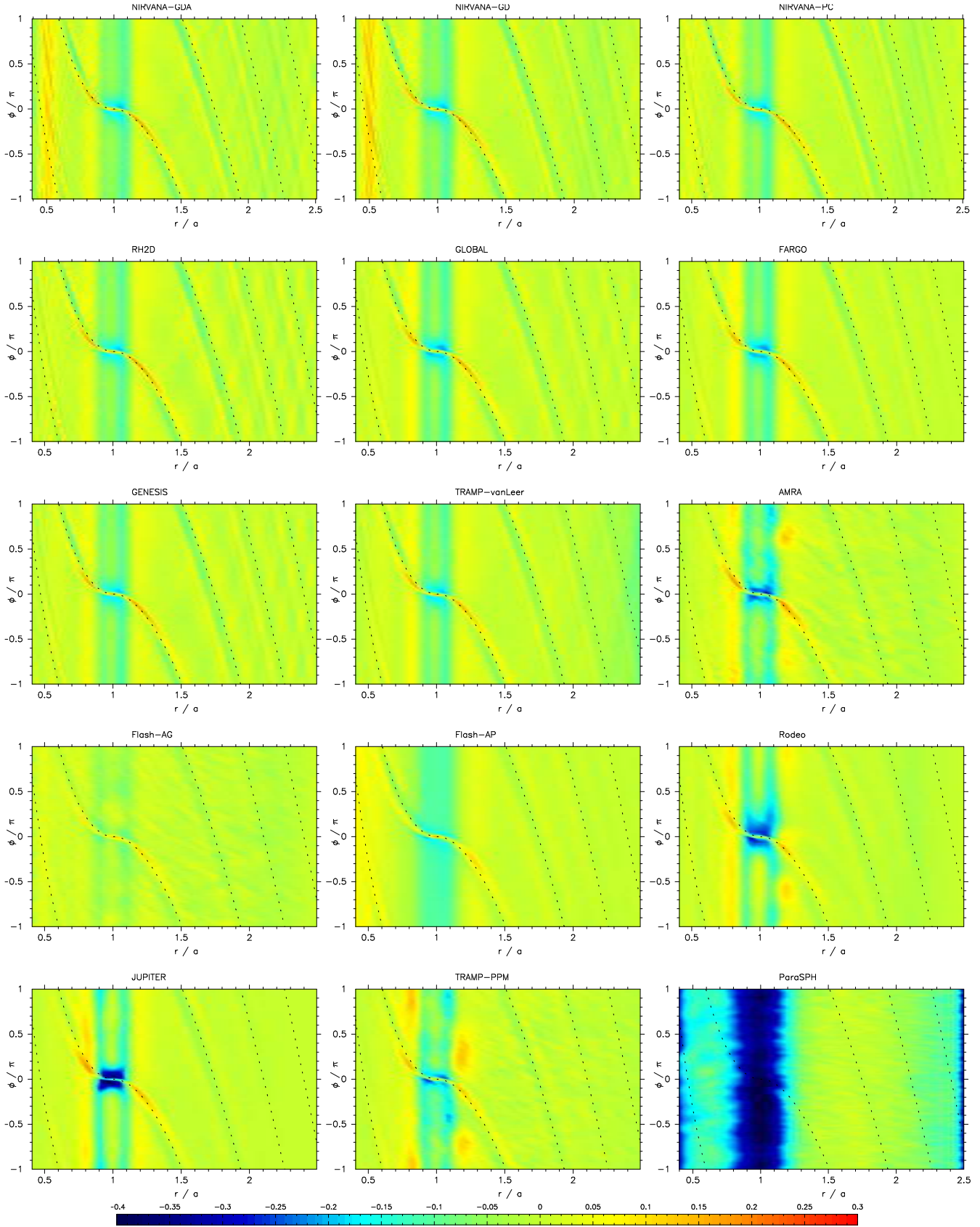

\includegraphics{figs/nep1.dens.100.ps}
\includegraphics{figs/nep.bar.dens.ps}
\caption{Surface density contours after 100 orbits
for the inviscid Neptune simulations.
The theoretical estimation of the spiral wakes is
represented by the dashed line.
The density scale ranges between $-0.4 < \log(\Sigma/\Sigma_0) < 0.3$.}
\label{fig:nepinviscdens}
\end{figure*}

The dip opened by Neptune after 100 orbital periods is much
shallower than for the Jupiter case.
The surface density maps are plotted in 
Figure~\ref{fig:nepinviscdens} 
for a Neptune mass planet embedded in an inviscid disc.
The spiral arms created by the planet are significantly weaker
than in the Jupiter calculations and are in better agreement
with the theoretical prediction of the shock
positions 
shown by the dashed line.
In the SPH simulations the shocks are extremely weak.
There are no overdense regions around
the Lagrangian points inside the gap in any of the calculations
since the gap is not deep enough.
Along the edge of the gap there are several blobs
in the \pci{}, \sijme{} and \trampppm{} results,
which are smaller than in the inviscid Jupiter calculations.
The \flash{}, \pci{} and \jupiter{} codes show ripples
in the disc and the gap with lower amplitude than in the inviscid
Jupiter simulations.

\begin{figure*}
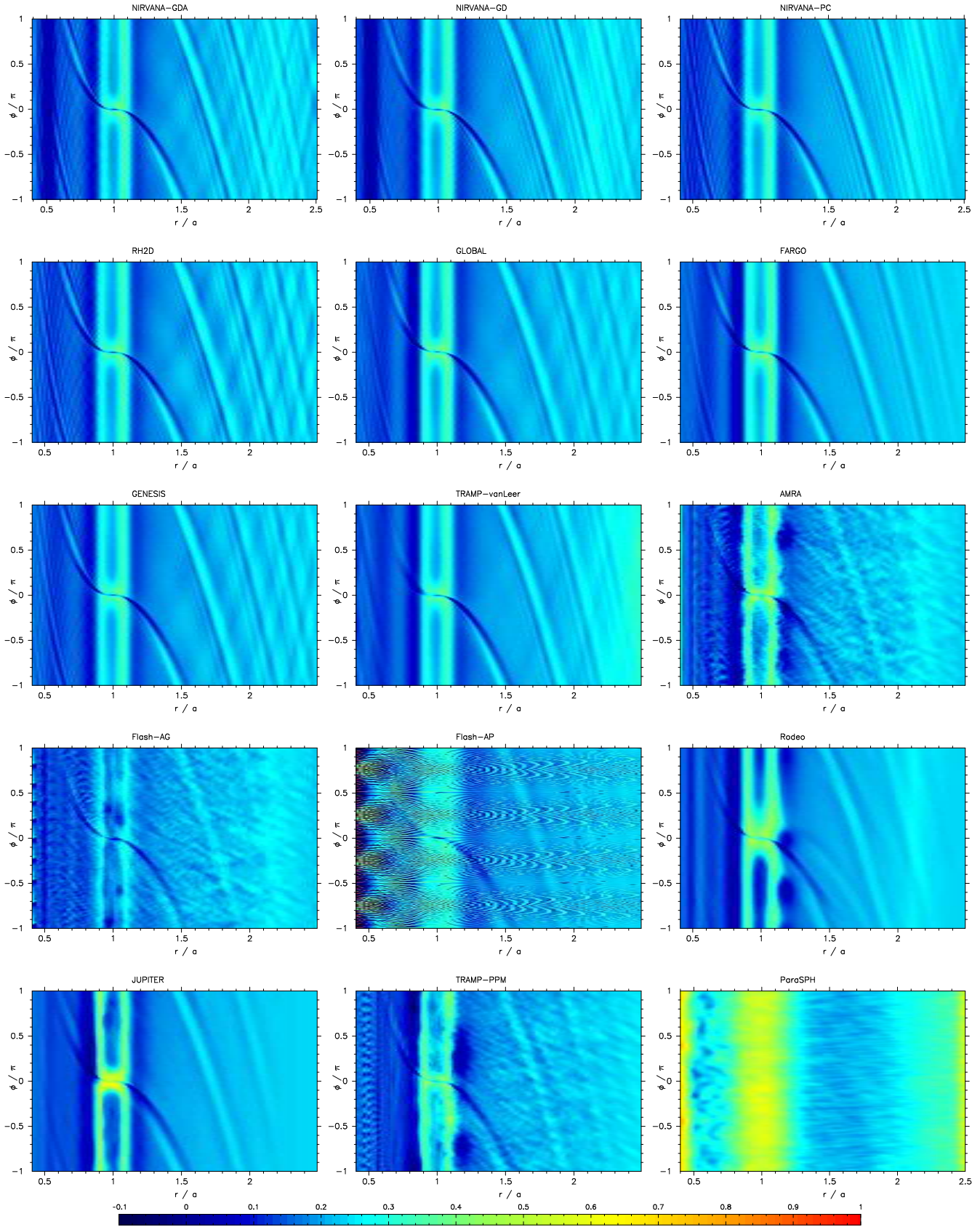

\includegraphics{figs/nep1.vort.100.ps}
\includegraphics{figs/nep.bar.vort.ps}
\caption{Vortensity contours in logarithmic scale
after 100 orbits for the inviscid Neptune calculations.
The vortensity range is $-0.1 < \log(\zeta) < 1$.}
\label{fig:nepinviscvort}
\end{figure*}

The comparative vortensity maps 
in the corotating frame
are shown in 
Figure~\ref{fig:nepinviscvort}.
Vortices moving along the gap are observed in the grid codes, although
they are smaller than in the Jupiter case. 

\begin{figure}
\includegraphics{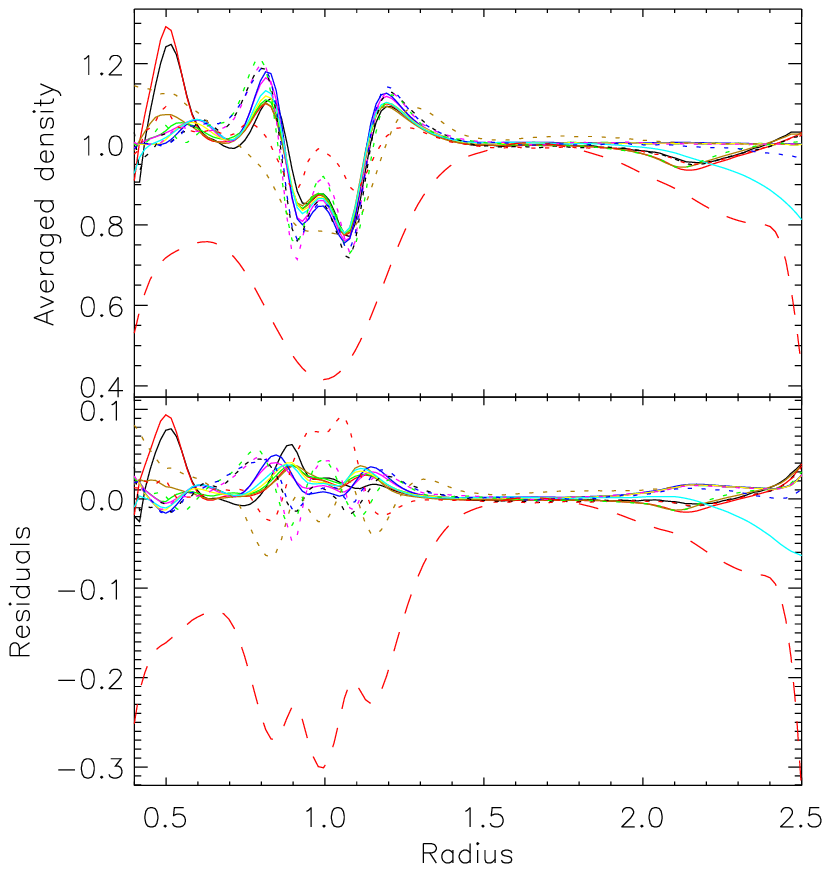}
\caption{Surface density profiles averaged azimuthally over $2\pi$
after 100 orbits for the inviscid Neptune runs are shown
in the upper panel.
The residuals in the lower panel are defined as in
Figure~\ref{fig:jupinviscprof}.}
\label{fig:nepinviscprof}
\end{figure}

\begin{figure}
\includegraphics{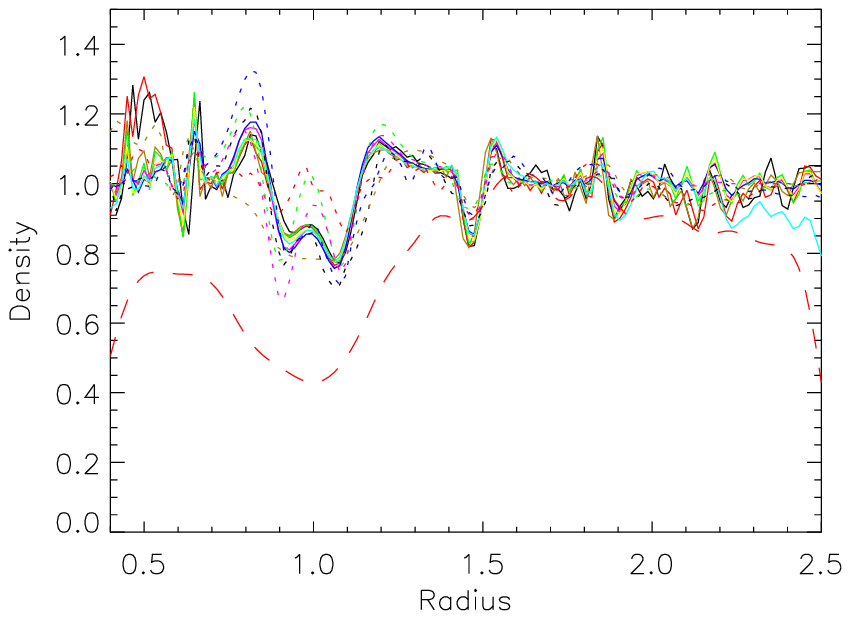}
\caption{Surface density profiles opposite to the planet position 
after 100 orbits for the inviscid Neptune runs.}
\label{fig:nepinviscprofpi}
\end{figure}

The azimuthally averaged density profiles 
after 100 orbits are plotted in Figure~\ref{fig:nepinviscprof}.
The depth and width of the gap is again in fairly good agreement
for the Eulerian codes.
\artur{}'s gap is shallower than the other grid based codes.
\parasph{} has a wider and deeper gap than the grid models
and a depleted inner disc.
The gap profile of the Eulerian codes is slightly asymmetric
with the deepest part just outside of the planet radius.
In the lower panel in Figure~\ref{fig:nepinviscprof}, we show
the residuals of the averaged profiles divided by the disc mass
with respect to the mean value.

In Figure~\ref{fig:nepinviscprofpi} we plot the 
surface density at $\phi=\pi$.
The shape and amplitude of the waves in the disc
agree well for the different codes outside the wave
damping boundaries.
There is a larger dispersion at the inner gap edge
and in the middle of the gap for the
\godunov\ schemes. The gap is slightly asymmetric
for the majority of the codes.


\begin{figure}
\includegraphics{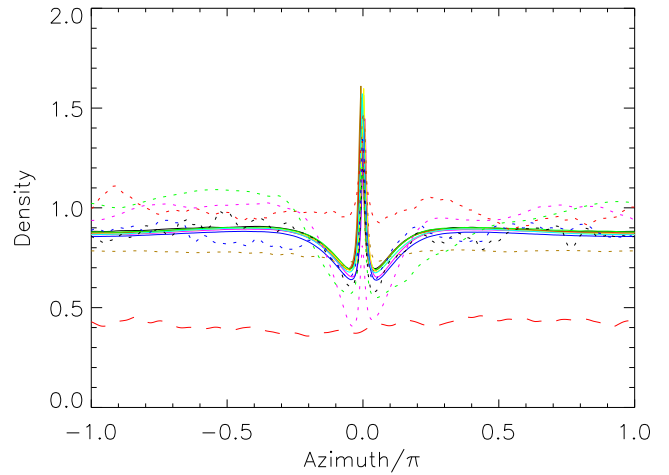}
\caption{Surface density azimuthal slice at the planet location 
after 100 orbits for the inviscid Neptune calculations.}
\label{fig:nepinviscphi}
\end{figure}

In Figure~\ref{fig:nepinviscphi} the azimuthal slices of 
surface density at the planet position are shown. 
A large density peak is observed again at the planet position
for all the grid codes.
The \artur{}, \sijme{} and \jupiter{} density
in the centre of the gap after 100 orbits
is close to the initial density
with depressions next to the planet. 
The rest of grid codes show a density decrease of
about 10-20\%. 
\parasph{} has the lowest density inside the gap with a 
decrease of about 60\% from the initial value. 


\begin{figure}
\includegraphics{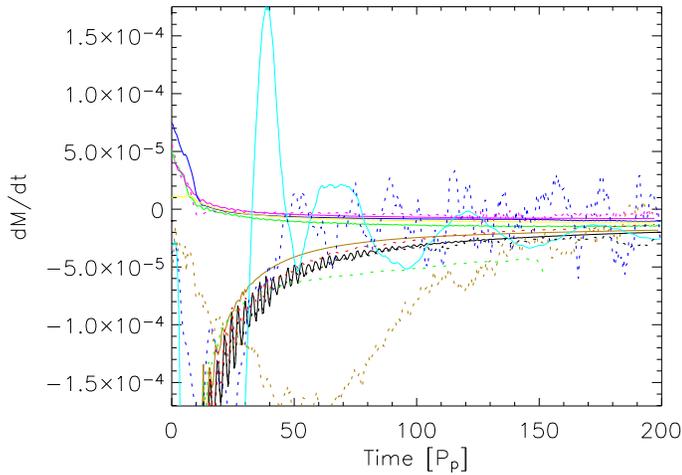}
\caption{Disc mass loss rate evolution for the inviscid Neptune 
simulations.}
\label{fig:nepinviscmass}
\end{figure}

Figure~\ref{fig:nepinviscmass} shows the grid mass loss rate as a
function of time for the grid-based codes.
All models show total mass loss due to the wave killing condition.
The \adam{} code has mass loss in the inner disc due to the
absence of a solid inner boundary in Cartesian coordinates but
converges to a value of a few times $10^{-5}$ after 200 orbits.
There is mass increase in the inner disc for some schemes in the
beginning of the simulation.
This suggests that there is gas flow trough the gap from the outer to 
the inner disc
since in the Neptune simulations the gap is shallower
and the planet generates weaker shocks.
The artificial viscosity
may cause the flow from the outer to the inner disc.
Another possible explanation is that the damping wave condition near
the inner boundary adds artificially angular momentum to the disc.


\begin{figure}
\includegraphics{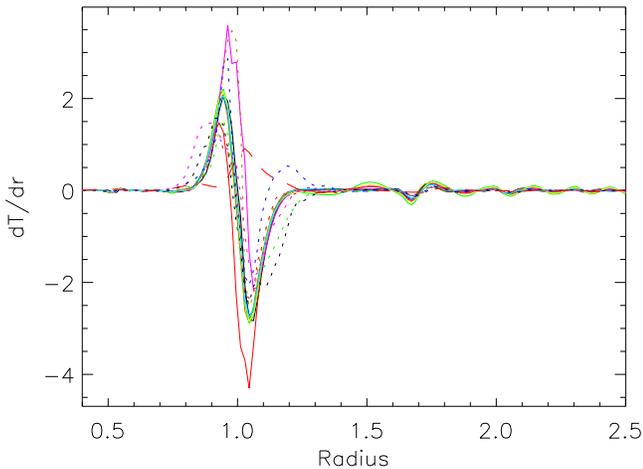}
\caption{Specific torque profiles after 100 orbits
for the inviscid Neptune simulations.}
\label{fig:nepinviscdTdr}
\end{figure}

We plot the profiles of
the derivative of the torque with respect to the radius
in Figure~\ref{fig:nepinviscdTdr}.
The agreement between the codes is good compared with
the inviscid Jupiter case, especially for the cylindrical grid
hydrodynamical codes.
The vortices do not perturb the planet strongly
and the specific torque radial profiles are stationary.
The outer disc generates again a negative torque acting on the planet
and the inner disc gives a positive torque.
At several Hill radii away from the planet location
the torques become negligible.

\begin{figure*}
\begin{tabular}{ccc}
\includegraphics{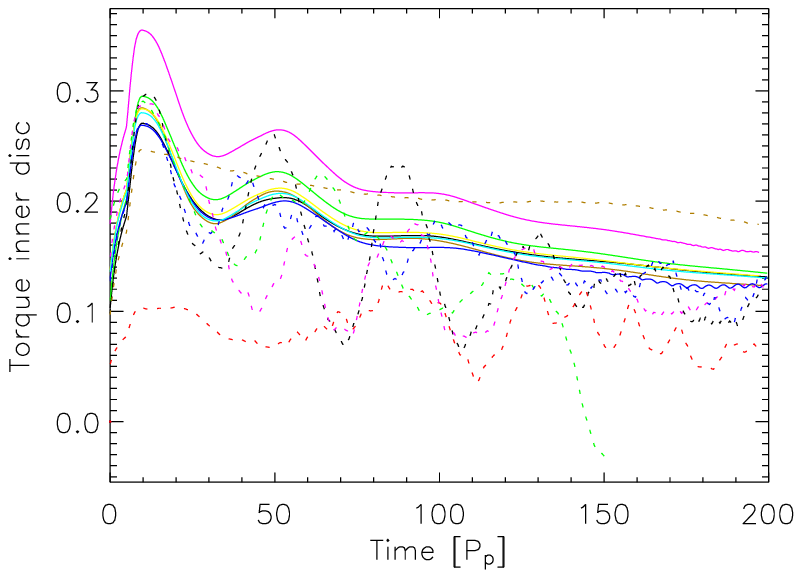} & \includegraphics{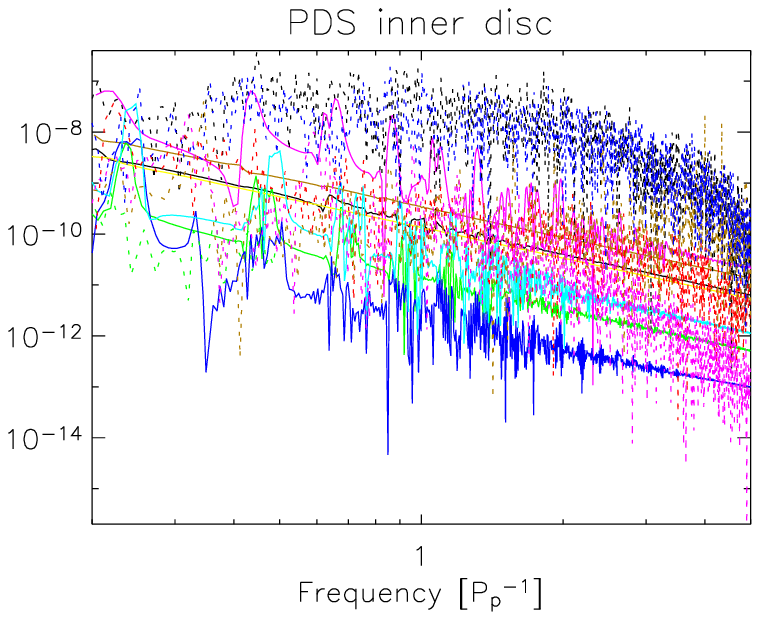} \\
\includegraphics{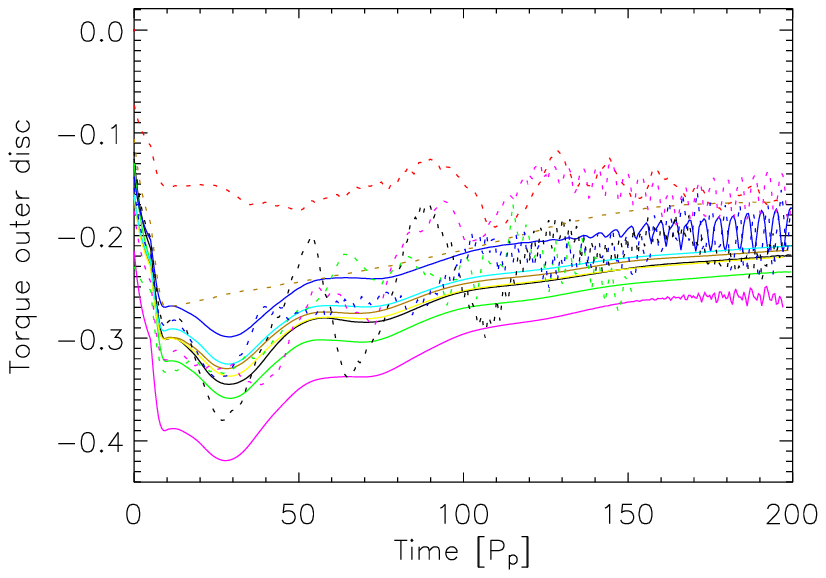} & \includegraphics{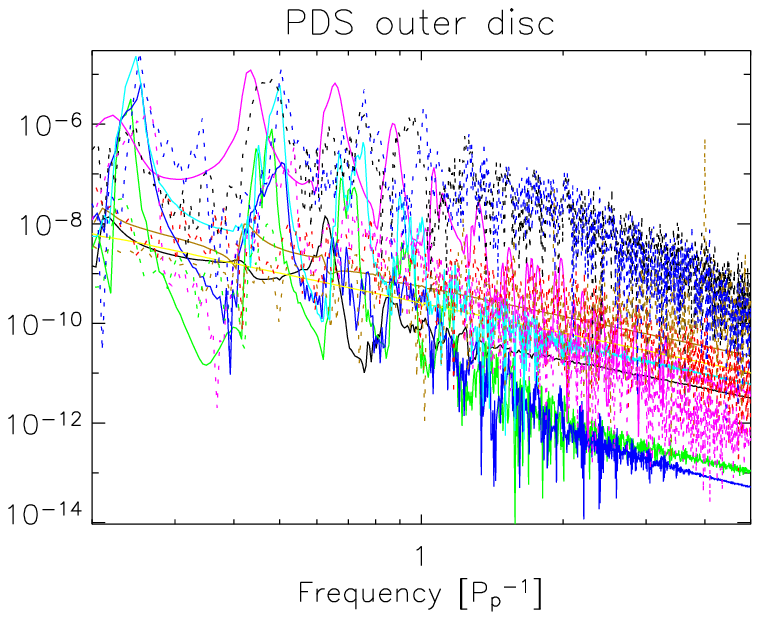} \\
\includegraphics{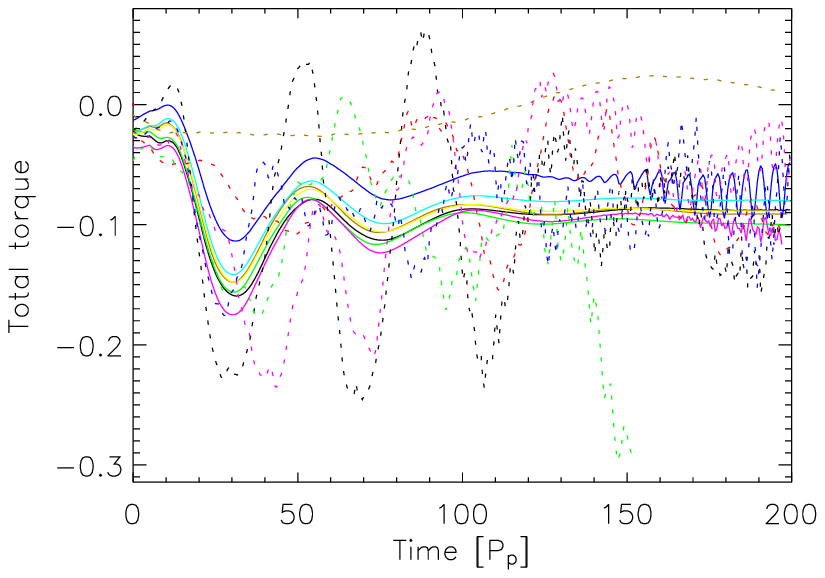} & \includegraphics{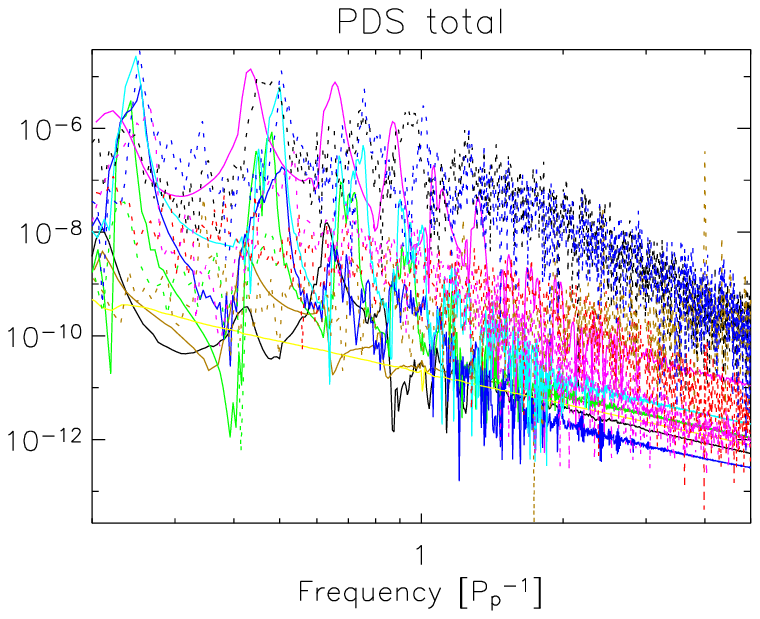}
\end{tabular}
\caption{Time averaged torques 
and their PDSs 
for the inviscid Neptune simulations.
The plots are shown in the same order as in Figure~\ref{fig:jupinviscpds}
and exclude the material inside the Roche lobe.}
\label{fig:nepinviscpds}
\end{figure*}

In Figure~\ref{fig:nepinviscpds},
the time average of the gravitational torques acting on the planet and
their associated PDSs are plotted.
The total torque after 200 periods agree within a factor 2
(see Table~\ref{tbl:torques.nep1}).
\upwind\ results show good agreement
while the \godunov\ results have larger oscillations.
The oscillations observed in the raw data and PDSs
may be produced by short-lived vortices
appearing during the first orbits which 
are not visible at later time in the density snapshots.


\begin{table}
\centering
\caption{Averaged torques
at the end of the simulations in units where
$a=1$, $P=2\pi$ and $M_{\text{*}}=1-\mu$
for the Neptune inviscid simulations.}
\begin{tabular}{lc}
\hline
Code & Torque \\
\hline
\input{torques.200.nep1.dat}
\hline
\end{tabular}
\label{tbl:torques.nep1}
\end{table}


\subsection{Viscous Neptune}

\begin{figure*}
\includegraphics{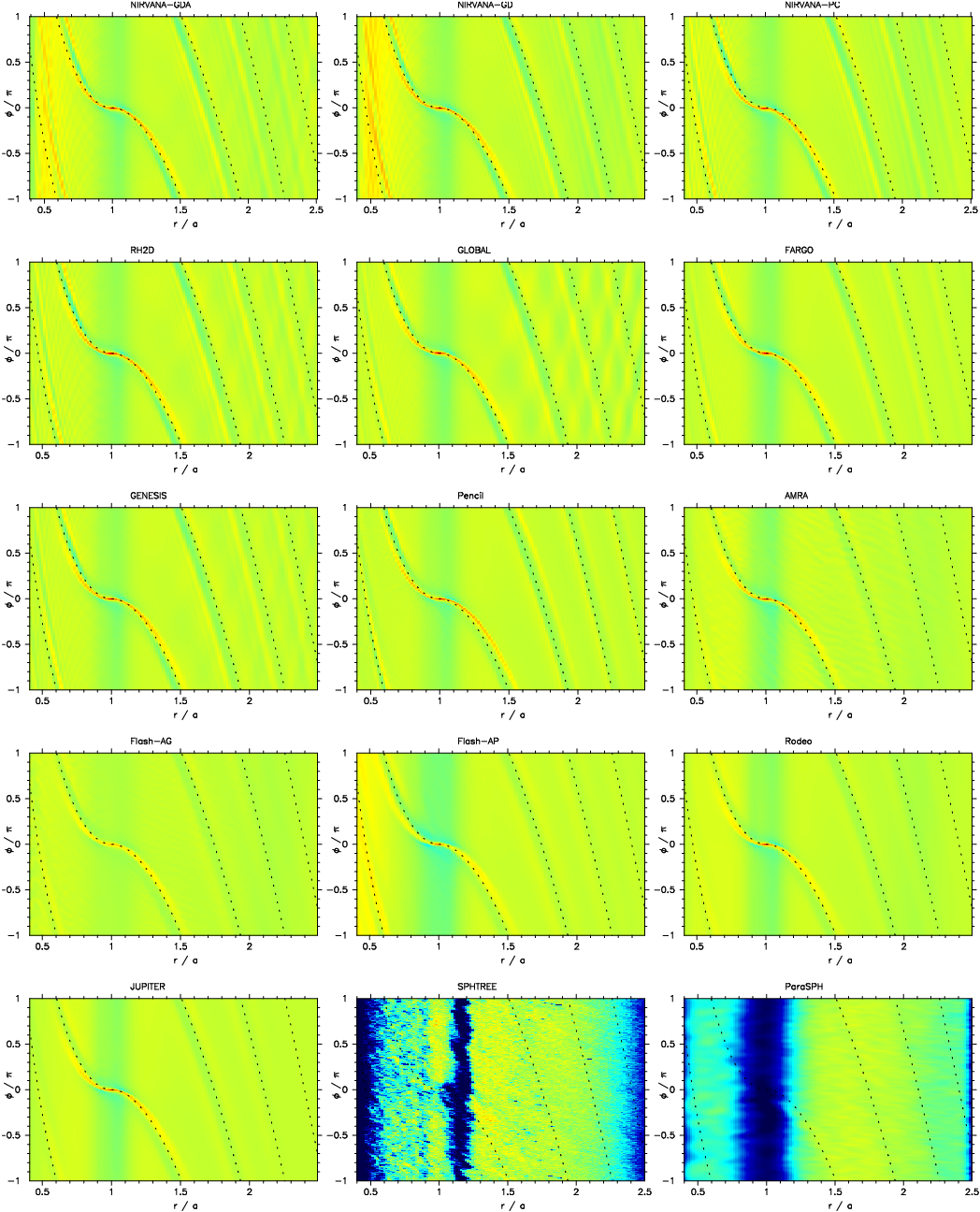}
\includegraphics{figs/nep.bar.dens.ps}
\caption{Surface density maps after 100 orbits
for the viscous Neptune simulations.
The dashed line is the estimated theoretical position
of the spiral arms.
The density range is $-0.4 < \log(\Sigma/\Sigma_0) < 0.3$.}
\label{fig:nepviscdens}
\end{figure*}

In Figure~\ref{fig:nepviscdens}
the comparative surface density contours after 100 orbits
for the viscous Neptune case are plotted.
The theoretical estimation of the spiral shocks positions
by \citet{2002MNRAS.330..950O} is shown by the dashed line.
The flow is smoother than in the inviscid Neptune simulations.
The density lumps moving along the edge of the gap
have disappeared and the planetary wakes are stable.
The filamentary structures in the \godunov\ simulations
have a reduced amplitude compared with the
inviscid case.

\begin{figure*}
\includegraphics{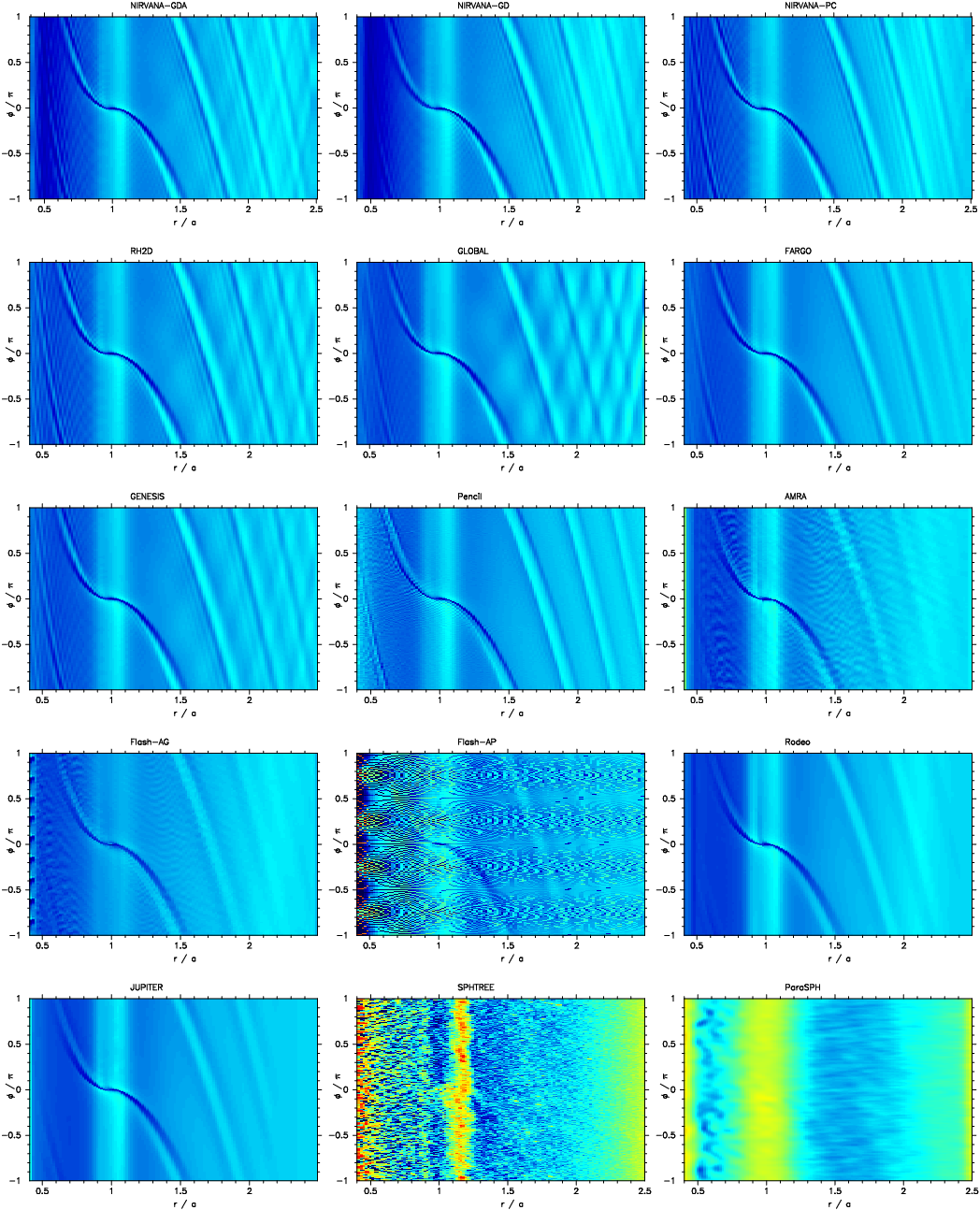}
\includegraphics{figs/nep.bar.vort.ps}
\caption{Vortensity contours after 100 orbits
for the viscous Neptune simulations.
The vortensity range is $-0.1 < \log(\zeta) < 1$.}
\label{fig:nepviscvort}
\end{figure*}

The vortensity maps are shown
in Figure~\ref{fig:nepviscvort}.
The density blobs lying next to the gap's edge
are not observed in the viscous simulations in logarithmic scale.
Several codes show wave reflection at the outer boundary despite
the wave damping condition.

\begin{figure}
\includegraphics{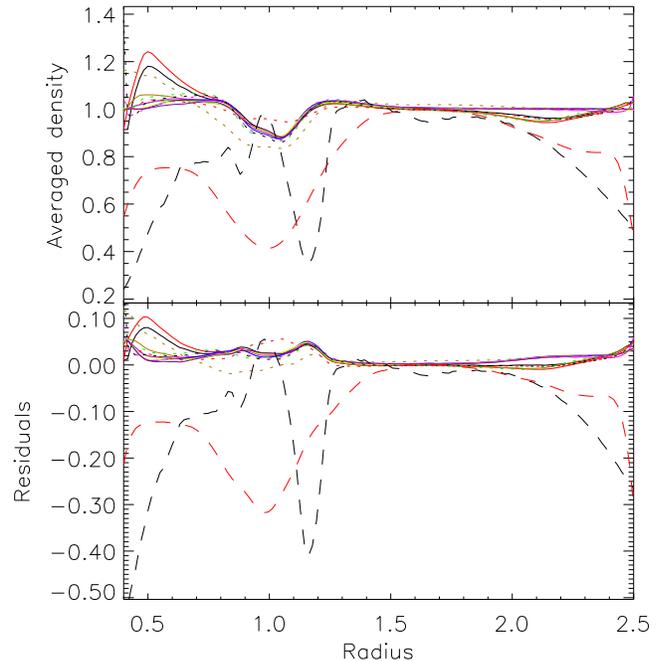}
\caption{The upper panel shows the surface density profiles averaged
azimuthally over $2\pi$
after 100 orbits for the viscous Neptune runs.
The residuals in the lower panel are defined as in
Figure~\ref{fig:jupinviscprof}.}
\label{fig:nepviscprof}
\end{figure}

\begin{figure}
\includegraphics{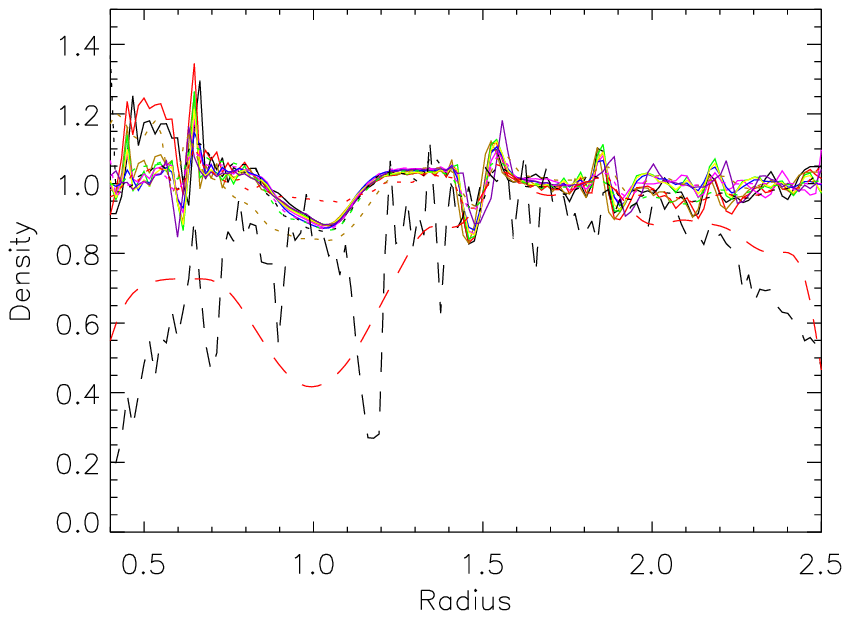}
\caption{Surface density profiles opposite to the planet position
after 100 orbits for the viscous Neptune case.}
\label{fig:nepviscprofpi}
\end{figure}

The smoothed density profiles are shown in
Figure~\ref{fig:nepviscprof}
for the viscous Neptune calculations.
The gap profile is again in good agreement for the polar grid
hydrodynamics codes.
The gap is shallower for \artur{} than for
the other Eulerian codes.
\adam{} has a wider and deeper gap with a flat shape.
\parasph{} has a very deep gap and \treesph{} has a strong
asymmetry with the deeper depression outside the planet postion.
The residuals of the averaged profiles divided by the disc mass
are shown in the bottom panel in Figure~\ref{fig:jupinviscprof}

The surface density opposite to the planet  
after 100 orbits is shown in Figure~\ref{fig:nepviscprofpi}.
\artur{} has a shallow gap whereas
\adam{} has a deeper and broader gap.
The waves observed in the inner and outer disc 
agree within a few percent for the Eulerian schemes.
\parasph{} has a very open gap and \treesph{} has a noisy profile
with a deep cavity outside the planet's radius.

\begin{figure}
\includegraphics{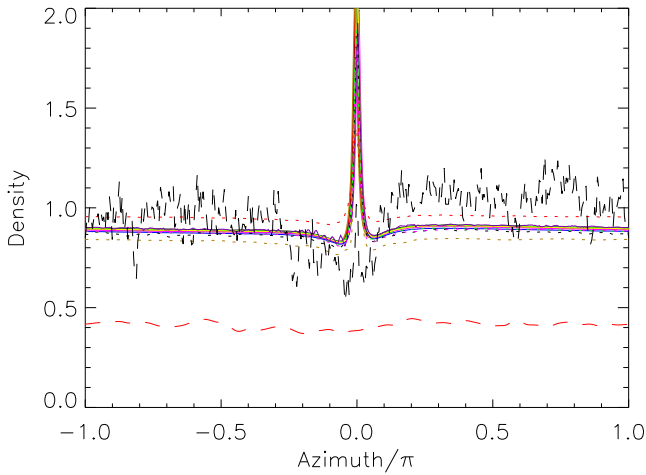}
\caption{Surface density azimuthal slice at the planet location 
after 100 orbits for viscous Neptune calculations.}
\label{fig:nepviscphi}
\end{figure}

The surface density azimuthal slices after 100 orbits are plotted in
Figure~\ref{fig:nepviscphi}.
A density spike appears again at the planet location 
in the grid codes.
The grid codes
show a density decrease of
approximately 10-20\% of the initial density in the center of the gap,
while \parasph{} has a  decrease of about 60\%
as in the inviscid Neptune case.


\begin{figure}
\includegraphics{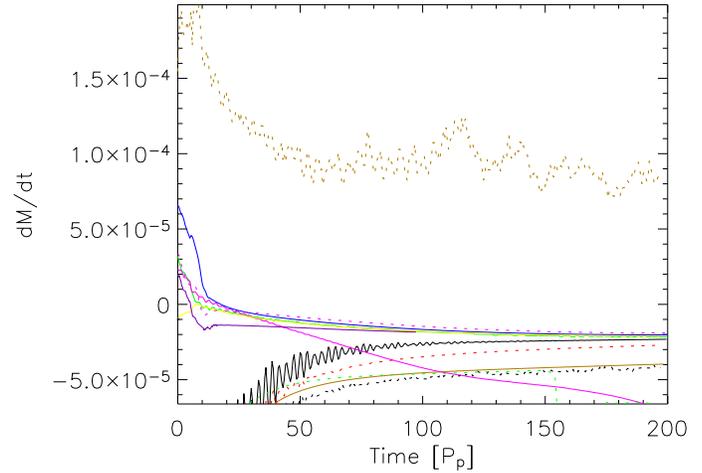}
\caption{Disc mass loss rate evolution for the viscous Neptune
simulations.}
\label{fig:nepviscmass}
\end{figure}

The disc mass loss rate is shown in Figure~\ref{fig:nepviscmass}.
All the models apart from \adam{} show total mass loss after 200 periods
with final values consistent within a factor of about 3.
\sijme{} has a sharp jump in mass loss rate at about 155 periods.
\adam{} results have a considerable mass
transfer from the region close to the star due to the gravitational softening.


\begin{figure}
\includegraphics{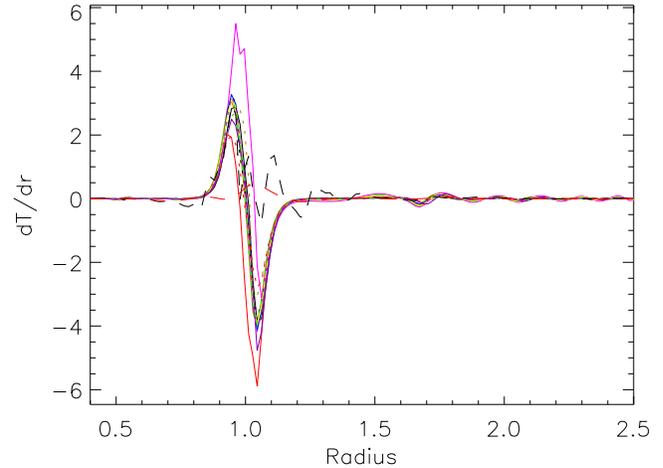}
\caption{Specific torque profiles after 100 orbits
for the viscous Neptune simulations.}
\label{fig:nepviscdTdr}
\end{figure}

The $dT/dr$ profiles after 100 orbits are shown in
Figure~\ref{fig:nepviscdTdr}.
The profiles show a good agreement between
the grid-based codes.

\begin{figure*}
\begin{tabular}{ccc}
\includegraphics{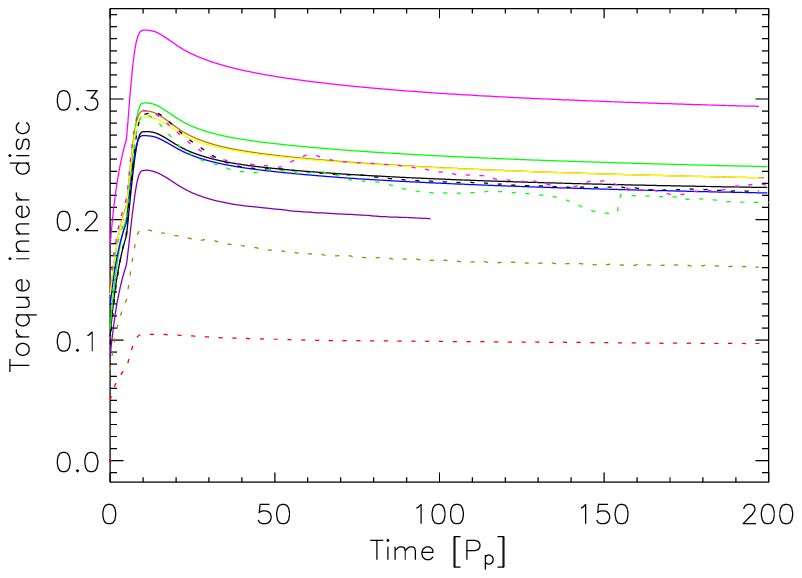} & \includegraphics{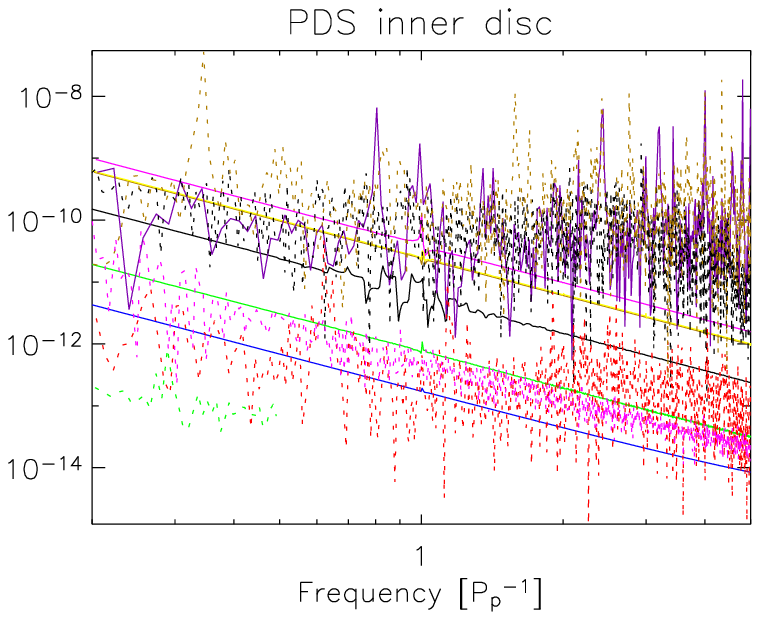} \\
\includegraphics{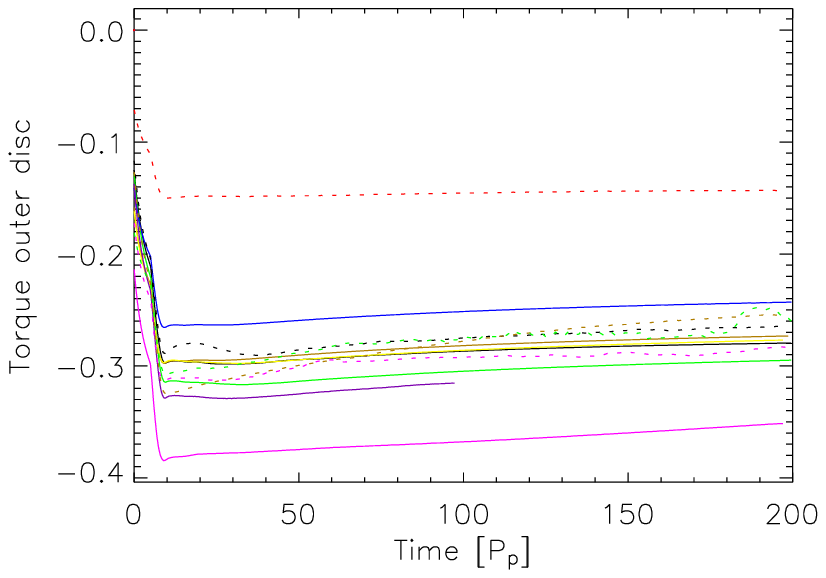} & \includegraphics{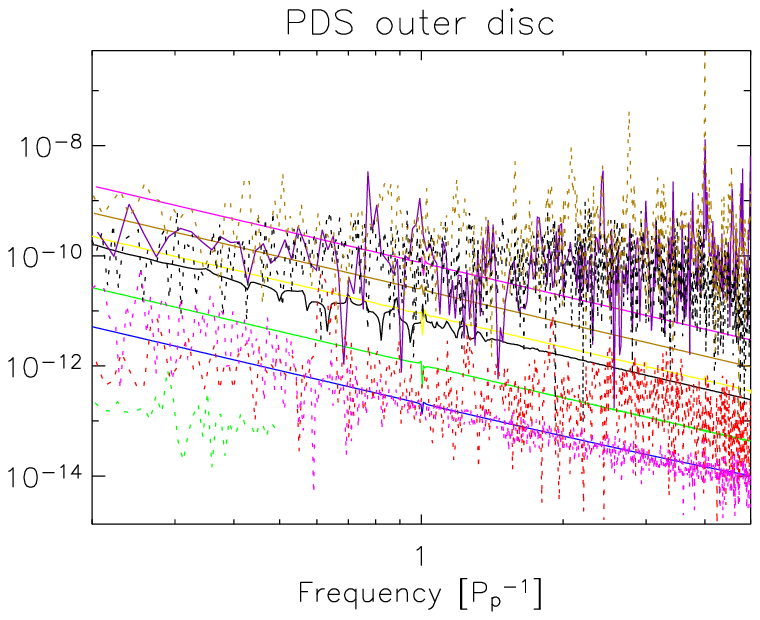} \\
\includegraphics{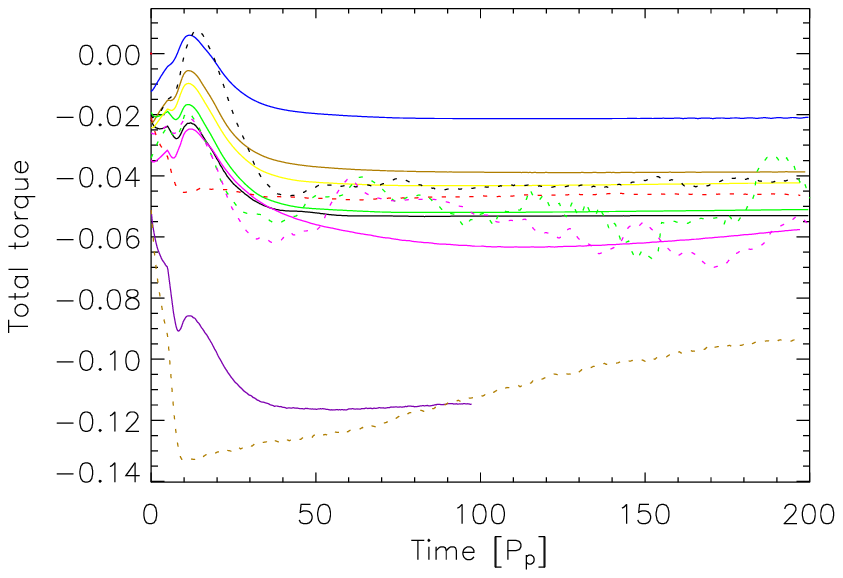} & \includegraphics{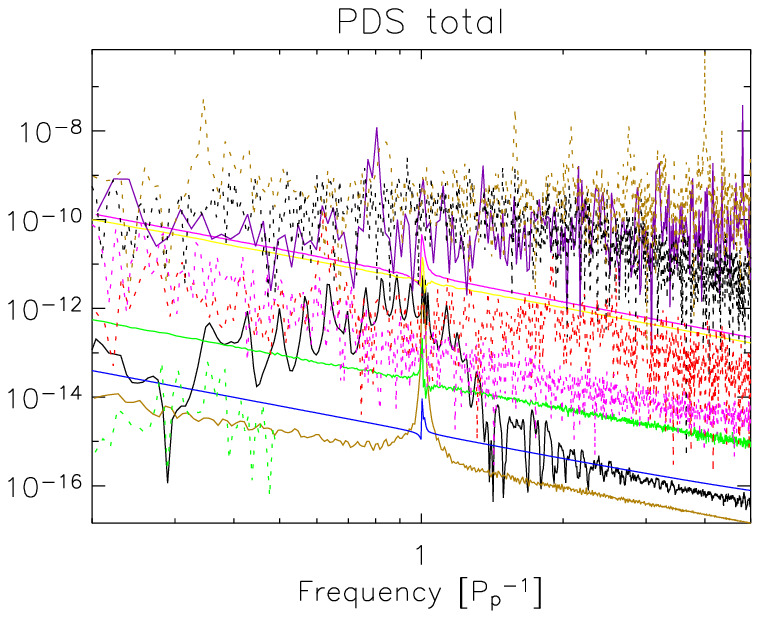}
\end{tabular}
\caption{Time averaged torques 
and corresponding PDSs 
for the viscous Neptune case.
The plots are shown in the same order as in Figure~\ref{fig:jupinviscpds}
and exclude the material inside the Roche lobe.}
\label{fig:nepviscpds}
\end{figure*}

We plot the time averaged torques acting on the planet
on the left hand side of Figure~\ref{fig:nepviscpds}.
The torque contribution from the inner disc
is positive, while the outer disc contribution is negative.
The outer disc dominates the total torque
and would cause an inwards orbital shift for a free moving planet.
In Table~\ref{tbl:torques.nep2}, we show
the averaged torques at 200 orbital periods.
The torques PDSs are shown
on the right hand side of Figure~\ref{fig:nepviscpds}.
The spectrum is rather flat for all codes which agrees with the
absence of vortices or eccentricity in the disc.

\begin{table}
\centering
\caption{Averaged torques in the window 175--200 periods
in units where
$a=1$, $P=2\pi$ and $M_{\text{*}}=1-\mu$
for the Neptune viscous simulations.}
\begin{tabular}{lc}
\hline
Code & Torque \\
\hline
\input{torques.200.nep2.dat}
\hline
\end{tabular}
\label{tbl:torques.nep2}
\end{table}

\subsection{High resolution simulations}
\label{sec:highres}

We studied the convergence of the results
running the test problem at 2 and 4 times the
original linear resolution with some of the codes.
\cresswell{} and \gerben{} Jupiter simulations
were run at resolution
$n_{\text{r}} \times n_{\phi}=(256,768)$.
Several tests at
$n_{\text{r}} \times n_{\phi}=(512,1536)$
were done with \kley{}, \gerben{} and \fargo{} codes
for Jupiter and Neptune planet masses.
\parasph{} was run using $853\,280$ particles and $146\,720$ boundary particles
for the Jupiter viscous case.

In the grid-based schemes,
the flow is observed to be smoother and more
stable in time than in the low resolution runs.
Vortices are still visible
in the Jupiter inviscid simulations
in the \cresswell{} and \fargo{} simulations. 
The vortices are more extended than in the
lower resolution calculations and
interact with the primary and secondary shocks.
There is more mass piling up inside the Roche lobe after 200 orbits in
the higher resolution cases in agreement with the results of
\citet{2005MNRAS.358..316D}.
Nevertheless, the averaged density profiles are very similar to the
results presented in the previous sections.
The gravitational torques are similar in the grid-based codes
and in good agreement with the low resolution results.

The \parasph{} results with $\sim 850\,000$ particles have stronger
shocks and the density profiles are in good agreement with
the grid-based results. This suggests that SPH schemes need higher
resolution to model accurately the corotation region and planetary wakes.

%% file: torques.200.jup1.dat
\dangelo{} & $-1.452354 \times 10^{-5}$ \\
\cresswell{} & $-1.512417 \times 10^{-5}$ \\
\kley{} & $-1.930871 \times 10^{-5}$ \\
\fromang{} & $-1.550768 \times 10^{-5}$ \\
\pierens{} & $-1.565123 \times 10^{-5}$ \\
\trampvanleer{} & $-1.818716 \times 10^{-5}$ \\
\pci{} & $-3.769203 \times 10^{-5}$ \\
\artur{} & $-7.014221 \times 10^{-5}$ \\
\adam{} & $-2.187462 \times 10^{-5}$ \\
\sijme{} & $-1.880762 \times 10^{-5}$ \\
\jupiter{} & $-2.721250 \times 10^{-5}$ \\
\trampppm{} & $5.611237 \times 10^{-6}$ \\

%% file: torques.200.jup2.dat
\dangelo{} & $-7.279010 \times 10^{-5}$ \\
\cresswell{} & $-8.591188 \times 10^{-5}$ \\
\kley{} & $-7.895477 \times 10^{-5}$ \\
\fromang{} & $-5.424962 \times 10^{-5}$ \\
\pierens{} & $-7.980679 \times 10^{-5}$ \\
\pencil{} & $-1.265820 \times 10^{-4}$ \\
\pci{} & $-7.269365 \times 10^{-5}$ \\
\artur{} & $-9.288739 \times 10^{-5}$ \\
\adam{} & $-6.681127 \times 10^{-5}$ \\
\sijme{} & $-1.061018 \times 10^{-4}$ \\
\jupiter{} & $-8.067247 \times 10^{-5}$ \\

%% file: torques.200.nep1.dat
\dangelo{} & $-5.601425 \times 10^{-5}$ \\
\cresswell{} & $-5.802065 \times 10^{-5}$ \\
\kley{} & $-6.329492 \times 10^{-5}$ \\
\fromang{} & $-6.539345 \times 10^{-5}$ \\
\pierens{} & $-5.645140 \times 10^{-5}$ \\
\trampvanleer{} & $-5.078578 \times 10^{-5}$ \\
\pci{} & $-8.154881 \times 10^{-5}$ \\
\artur{} & $-6.523340 \times 10^{-5}$ \\
\adam{} & $1.055228 \times 10^{-5}$ \\
\sijme{} & $-1.234136 \times 10^{-4}$ \\
\jupiter{} & $-2.834292 \times 10^{-5}$ \\
\trampppm{} & $-5.742200 \times 10^{-5}$ \\

%% file: torques.200.nep2.dat
\dangelo{} & $-3.375560 \times 10^{-5}$ \\
\cresswell{} & $-2.467831 \times 10^{-5}$ \\
\kley{} & $-3.257739 \times 10^{-5}$ \\
\fromang{} & $-3.742940 \times 10^{-5}$ \\
\pierens{} & $-2.700463 \times 10^{-5}$ \\
\pencil{} & $-7.303624 \times 10^{-5}$ \\
\pci{} & $-2.634983 \times 10^{-5}$ \\
\artur{} & $-2.933768 \times 10^{-5}$ \\
\adam{} & $-6.022857 \times 10^{-5}$ \\
\sijme{} & $-2.697915 \times 10^{-5}$ \\
\jupiter{} & $-3.809917 \times 10^{-5}$ \\

%% file: discussion.tex

\section{Discussion}
\label{sec:discussion}

In this paper, we have studied a planet in a fixed orbit
embedded in a disc using 17 different SPH and Eulerian
methods.
The codes used in the comparison have been thoroughly
tested in problems with known analytical solutions.
The goal of this project was to investigate the reliability
of current astrophysical hydrodynamic codes in the disc-planet problem,
and to provide a reference for future
calculations. Performing this comparison also aided in
the debugging of the codes.

The results show good agreement on the general picture,
although there are some differences in the details.
The density maps and averaged profiles are consistent for
the grid-based methods.
The variation in the disc mass is of the order of 10\%
after 100 orbital periods, but this
does not seem to produce big differences in the surface
density distributions.
The different boundary conditions tested in \fargo{}
do not affect the results
since the goal in both boundary
implementations was to avoid the reflection of waves.
A preliminary study of convergence using finer grids shows
that there is agreement at 2 and 4 times the original
linear resolution.

Vortices are visible in the inviscid runs for both planet masses
$\mu = 10^{-3}$ and $10^{-4}$ in the grid codes, which
induce a strong perturbation to the tidal torque.
The vortices in the \upwind\ simulations have a larger amplitude
and are more extended than in the \godunov\ results. 
The total torque acting on the planet excluding the material
inside the Roche lobe
agree in order of magnitude
for Jupiter models.
The torque results for Neptune have greater dispersion,
possibly due to incomplete clearing of the gap,
but agree nevertheless in the final value within a factor 2.


It has been observed that \godunov\ codes show a large amount of
filamentary small-scale structure unseen in model results obtained
with other codes. This is especially true for both Direct-Eulerian
implementations, \pci{} and \flash{}. In addition, \pci{} results show
enhanced small-scale structure when compared to \flash{}. Extensive
comparison tests of the two implementations has shown that much of the
observed differences is due to use of more selective dissipation
algorithm in \pci{}. (The so-called flattening algorithm in \pci{} is
based on Eqns. A.7-A.10 from
\citet{1984JCP...54..174} while \flash{}
uses Eq. A.2). After adopting the simplified version of the flattening
algorithm in \pci{}, the results closely matched those obtained with the
\flash{} code. Adding a small amount of artificial viscosity with
coefficient of 0.1, as recommended by \citet{1984JCP...54..174}, 
resulted in a further reduction of filamentary structures and
substantial reduction of the strength of vortices located at the gap
edges.

The \upwind\ codes have
a smooth disc structure and do not show
filaments in the inviscid simulations.
This may be due to the fact that \godunov\ codes have small intrinsic
viscosity 
in our problem in cylindrical coordinates,
where flow is dominated by advection in only one dimension.
\citet{1999ApJ...514..344B} have shown that van Leer based codes
in polar coordinates may have low intrinsic viscosity comparable
with \godunov\ methods.
It has been checked that none of the above changes are needed
in \pci{} if the grid resolution is increased twice. In this case the
solution is much smoother and the vortices at the gap edges decay faster.

The Cartesian implementations produce results that are comparable
to the other codes but there are differences in the
gap structure due to the open inner boundary.
The depleted density distribution in the inner disc in \adam{} produces
different torques but the torque contribution from the outer disc is
consistent with the cylindrical grid codes.

SPH codes predict the shape of the gap correctly
but do not resolve well low density regions where the
number of particles is small.
In addition the spiral wakes are weaker,
possibly due to SPH being more dissipative.
The Balsara switch included in the \treesph{} code
is used to reduce the shear component of artificial
viscosity but it may also smooth out the shocks.
An advantage of SPH codes is that the geometry of the problem
is well adapted to a Lagrangian scheme and the algorithm
implementation is simpler than for Eulerian codes.
The planet can be treated as a regular particle which accretes
material. Furthermore, it is possible to follow the trajectory
of individual fluid elements and study the accretion flows.
SPH codes are computationally more expensive than Eulerian
codes at the same resolution. Our high resolution tests
indicate that higher resolution is needed in the SPH simulations
to obtain results comparable to the Eulerian grid codes.

Possible future work includes the comparison
of high resolution runs using multi-level meshes to
investigate the gas flow close to the planet,
the study of
the orbital shift of a free-moving planet
and 3-dimensional simulations
\citep[see e.g.][]{2001ApJ...547..457K,2003ApJ...586..540D}.
The convergence of the results with resolution needs to be
studied in detail.

In closing we would like to reiterate that computational work might be
regarded as an experiment, rather than a simulation. We have shown that
different codes can give slightly different results for the same physical
problem. Reproducibility of experimental results is fundamental to the
scientific process, and this standard must be applied to those performed with
computers. Before a computational result can be regarded as reliable, it must
be confirmed by an independent test with a different code.

%% file: advice.tex

\appendix
\section{Advice for Others}
\label{sec:appendix}

Putting together this comparison has been something of a ``learning experience'' for all concerned.
Although not strictly scientific, we would like to share our experiences with others who may be contemplating similar comparisons.

As with many things, advance planning is the most important.
So far is possible, decide in advance which quantities should be monitored, and how often this should be done.
What should be checked every timestep (or so), and what is only required at much less frequent intervals?
Storage requirements are relevant to this: for example, writing out the total mass in the simulation is a lot cheaper (both in terms of space and time) than outputting the entire density field.
Changing the output quantities at a later date will often involve re-running computations, which will delay matters.

Careful attention should also be paid to the format of the submitted files.
Each code generally has its own output format.
Those co-ordinating the comparison do not have time to pick through each of these - automated processing \emph{must} be the goal.
Make sure that the format is carefully specified (since if there are two mutually incompatible ways of doing something, it is certain that results with both ways will be submitted).
As an aside, for grid based results, it is probably more sensible to write out the indices of each cell, rather than the co-ordinates themselves: integers are exact.
Supply the tables to convert indices to co-ordinates separately.

Pay similar attention to the problem specification itself.
Some flexibility will inevitably be needed, but try to keep this to a minimum.
Again, the authors' experience is that anything left vague will be done in different ways by different groups.

Communication is also hugely important.
In addition to setting up a mailing list, the authors were able to hold several short meetings, using funds provided by the EU.
These were crucial to moving the project forward.
Better still would have been to have held a longer workshop (perhaps a week) where everyone could gather, discuss and run their codes together.

We hope that future groups will find our experiences useful in planning their own code comparisons.